\documentclass[fleqn,10pt]{wlscirep}
\usepackage[utf8]{inputenc}
\usepackage[T1]{fontenc}
\usepackage{bm}
\usepackage{graphicx}

\title{Extreme light confinement and control in low-symmetry phonon-polaritonic crystals} %AKA Fantastic polaritons and where to find them :)

\author[1]{Emanuele Galiffi$^\dagger$}
\author[2]{Giulia Carini$^\dagger$}
\author[3]{Xiang Ni$^\dagger$}
\author[4,5]{Gonzalo Álvarez-Pérez}
\author[1]{Simon Yves}
\author[1,6]{Enrico Maria Renzi}
\author[1]{Ryan Nolen}
\author[2]{Sören Wasserroth}
\author[2]{Martin Wolf}
\author[4,5]{Pablo Alonso-Gonzalez}
\author[2]{Alexander Paarmann*}
\author[1,6]{Andrea Alù**}
\affil[1]{Advanced Science Research Center, City University of New York, 85 St. Nicholas Terrace, 10031 New York, NY, USA}
\affil[2]{Fritz Haber Institute of the Max Planck Society, Berlin, Germany}
\affil[3]{School of Physics and Electronics, Central South University, Changsha, Hunan 410083, China}
\affil[4]{Department of Physics, University of Oviedo, 33006, Oviedo, Spain}
\affil[5]{Center of Research on Nanomaterials and Nanotechnology, CINN (CSIC-Universidad de Oviedo), El Entrego, Spain}
\affil[6]{Physics Program, The Graduate Center, City University of New York, 10026 New York, NY, USA}

\affil[*]{e-mail: alexander.paarmann@fhi-berlin.mpg.de}
\affil[**]{e-mail: aalu@gc.cuny.edu}
\affil[$\dagger$]{These authors contributed equally to this work.}

\begin{abstract}
Polaritons are a hybrid class of quasiparticles originating from the strong and resonant coupling between light and matter excitations. Recent years have witnessed a surge of interest in novel polariton types, arising from directional, long-lived material resonances, and leading to extreme optical anisotropy that enables novel regimes of nanoscale, highly confined light propagation. While such exotic propagation features may also be in principle achieved using carefully designed metamaterials, it has been recently realized that they can naturally emerge when coupling infrared light to directional lattice vibrations, i.e., phonons, in polar crystals. Interestingly, a reduction in crystal symmetry increases the directionality of optical phonons and the resulting anisotropy of the response, which in turn enables new polaritonic phenomena, such as hyperbolic polaritons with highly directional propagation, ghost polaritons with complex-valued wave vectors, and shear polaritons with strongly asymmetric propagation features. In this Review, we develop a critical overview of recent advances in the discovery of phonon polaritons in low-symmetry crystals, highlighting the role of broken symmetries in dictating the polariton response and associated nanoscale-light propagation features. We also discuss emerging opportunities for polaritons in lower-symmetry materials and metamaterials, with connections to topological physics and the possibility of leveraging anisotropic nonlinearities and optical pumping to further control their nanoscale response.
\end{abstract}
\begin{document}

\flushbottom
\maketitle

% \noindent For chemical structures, please see the \href {https://www.nature.com/documents/nr-chemical-structures-guide.pdf}{Nature guide for ChemDraw structures}.\\

% \noindent For information about the artwork process at Nature Reviews journals, please refer to \href{https://www.nature.com/reviews/pdf/artworkguide.pdf}{the guide to authors} for figures and an  \href{http://www.nature.com/reviews/pdf/artworkguidep2.pdf}{example figure}.\\

\thispagestyle{empty}

%\section*{Main text}
% \textit{Total text of article 6,000–8,000 words (i.e. from the Introduction to the Outlook section; without including the abstract, figure captions or text boxes).}
\section*{Introduction}
\label{sec:intro}

% \textit{We recommend that you keep the introduction to 800 words maximum. The introduction should answer the following questions: what is the topic of the article? Why is this topic of importance? What recent developments make it timely to review this topic now? And how will this topic be reviewed? To answer the last question, we recommend adding a guiding paragraph at the end of your introduction that clearly states what will (and what won’t) be discussed in your Review, so readers and referees will know what to expect.}

Polaritons are hybrid light-matter excitations, emerging from the strong coupling between photons and dipolar material excitations. The first-observed form of polariton, the plasmon polariton~\cite{wood1902xlii,ritchie1957plasma}, emerges from the coupling of photons with the oscillations of the quasi-free electron plasma in metals, leading to highly confined optical fields and largely enhanced light-matter interactions. The past two decades have witnessed tremendous research efforts in the associated field of plasmonics~\cite{maier2007plasmonics}, which leverages plasmon polaritons to unveil  nanophotonic technologies that exploit their features. In parallel, the past few years have experienced a broadening of polariton research, as polaritonic coupling based on several other material excitations, such as phonons, excitons and Cooper pairs, has gained considerable attention~\cite{basov2016polaritonreview,low2017polaritons,basov2021polaritonpanorama}. As a result, we are currently witnessing different branches of polariton research which, besides nanophotonics, address several aspects of light-matter coupling, such as polariton-driven chemistry~\cite{ribeiro2018polaritonchemistry}, vibrational strong coupling~\cite{thomas2019vibrationalstrongcoupling}, quantum optics~\cite{herrera2020polaritonquantumoptics}, ultra- and deep- strong coupling~\cite{mueller2020deepstrong}, and exciton-polariton physics in 2D materials~\cite{huang2022enhanced}. For nanophotonic applications at infrared frequencies, phonon polaritons~\cite{Foteinopoulou2019PhononPolaritons,gubbin2022surfacephononpolaritons} (PhPs) have gained significant attention for two features: (i) the low optical losses of PhPs in contrast to other polariton families~\cite{caldwell2015low,khurgin2018phononplasmon}, and (ii) the inherently anisotropic response of PhPs~\cite{he2022anisotropy}, due to the directionality of phonon oscillators in many polar crystals, associated with their non-trivial structural symmetry. In the following, we higlight and discuss the role of directionality and broken symmetry for polaritonics, focusing on the role of crystal symmetry in establishing anisotropic PhPs in natural materials, and discussing the opportunities that arise in these low-symmetry materials for polariton physics and applications. 

\begin{figure}[ht!]
    \centering
    \includegraphics[width=\textwidth]{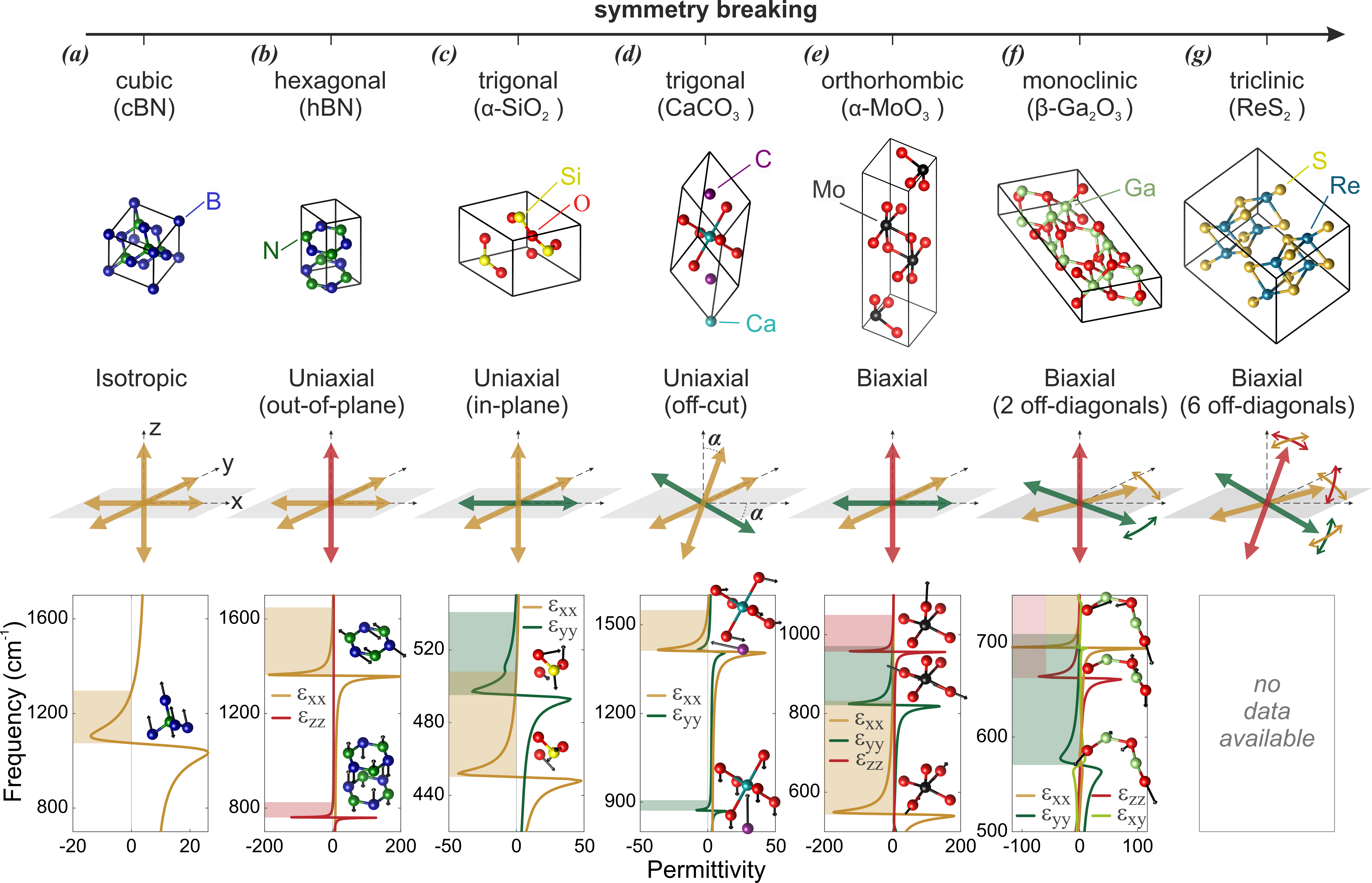}
    \caption{\textbf{Structural and optical symmetry breaking in polar crystals}. From left to right, a reduction in the polar crystal structural symmetry (top row) is associated with a corresponding symmetry reduction in the optical response (middle row), enhanced by sharp and directional phonon resonances in the infrared permittivity response (bottom row), where we also schematically show the corresponding vibrational mode patterns\cite{Jain2013materialsproject}. Departing from \textbf{(a)} cubic crystals, such as cubic boron nitride (cBN) with isotropic optical response supporting SPhPs, we move to \textbf{(b)} hexagonal and \textbf{(c,d)} trigonal crystals with uniaxial optical properties. Out-of-plane uniaxial hBN \textbf{(b)} supports ultra-confined hyperbolic PhPs\cite{caldwell2014sub,dai2014tunable,li2015hyperbolic}, while \textbf{(c)} in-plane uniaxial $\alpha$-SiO$_2$ and \textbf{(d)} off-cut calcite (CaCO$_3$) support highly directional in-plane hyperbolic surface polaritons~\cite{passler2022hyperbolic} and ghost hyperbolic polaritons~\cite{ma2021ghost}, respectively. Directional leaky polaritons have been observed in in-plane and off-cut uniaxial crystals~\cite{ni2023observation}. \textbf{(e)} Orthorhombic 2D crystals, such as $\alpha$-MoO$_3$, show biaxial responses, which enable ultra-confined in-plane hyperbolic modes~\cite{ma2018plane}. \textbf{(f)} Monoclinic crystals, like $\beta$-Ga$_2$O$_3$, show biaxial responses with frequency-dependent OA orientation and a non-diagonal permittivity tensor, leading to the formation of hyperbolic shear polaritons~\cite{passler2022hyperbolic,hu2023real}. \textbf{(g)} Even more exotic phenomena may be expected for triclinic crystals like ReS$_2$~\cite{mosshammer2022RS2triclinic}, but polariton observations in these materials are still missing. Notably, the combination of crystalline broken symmetry (top row) and sharp directional resonances leads to dramatic anisotropy in the optical response (bottom row), facilitating the exotic polaritonic phenomena discussed in this Review.} 
    \label{fig:fig1}
\end{figure}

Bulk phonon polaritons were first discussed by Born and Huang~\cite{bornhuang1954} in 1954 and experimentally observed by Henry and Hopfield~\cite{henry1965raman}. Barker~\cite{barker1972direct} first measured the dispersion of surface phonon polaritons (SPhPs), while the effects of anisotropy were first explored by Falge and Otto~\cite{falge1973dispersion}, without however realizing its profound impact on nanoscale light propagation. The potential for coherent thermal emission from SPhPs was demonstrated employing SiC line gratings~\cite{greffet2002coherent}, while only recently sub-diffractional nanostructures supporting localized PhP resonances were realized~\cite{caldwell2013low}. The experimental capabilities of scattering-type near-field optical microscopy (s-SNOM)~\cite{huber2005near} opened vast opportunities for real-space imaging of propagating PhPs. A real breakthrough in polariton research followed the advent of two-dimensional (2D) van der Waals (VdW) materials~\cite{geim2007rise} and heterostructures~\cite{geim2013van}, initially focusing on graphene plasmonics~\cite{koppens2011graphene,chen2012optical,fei2012gate}. Research on PhPs in 2D materials sparked with the discovery of the extreme anisotropy of phonon resonances in the VdW crystal hexagonal boron nitride (hBN)~\cite{dai2014tunable,caldwell2014sub,li2015hyperbolic} (see Fig.~\ref{fig:fig1}b), supporting hyperbolic polaritons, which had so far only been observed in plasmonic metamaterials~\cite{liu2007far}. Extensive work on hBN initiated a cascade of new discoveries driven by further reducing the crystal symmetry, as illustrated in Fig.~\ref{fig:fig1}. Moving past hBN (Fig.~\ref{fig:fig1}b), uniaxial crystals such as $\alpha$-SiO$_2$ (Fig.~\ref{fig:fig1}c) or CaCO$_3$ (Fig.~\ref{fig:fig1}d) support similarly directional and sharp phonon resonances, but can be cut with arbitrary orientation of the optical axis (OA) with respect to their interfaces, leading to the emergence of in-plane hyperbolic SPhPs~\cite{passler2022hyperbolic}, and even more exotic responses, such as ghost hyperbolic polaritons~\cite{ma2021ghost}, and directional leaky polaritons~\cite{ni2023observation}, which leverage nontrivial rotations between the OA and the material interface. Biaxial 2D crystals, such as orthorhombic alpha-phase molybdenum trioxide ($\alpha$-MoO$_3$), support volume-confined in-plane hyperbolic PhPs~\cite{ma2018plane} (Fig.~\ref{fig:fig1}e), further broadening the potential for applications of PhPs. Recently, hyperbolic shear polaritons~\cite{passler2022hyperbolic} were discovered in lower-symmetry monoclinic crystals (Fig.~\ref{fig:fig1}f), associated with more directional and asymmetric propagation features. This synergy between lattice broken symmetries and long-lived phonon resonances, at the basis of this exciting range of phonon polaritons, holds the promise to even more exciting opportunities emerging in triclinic crystals (Fig.~\ref{fig:fig1}f), currently under investigation. In the following, we unveil this path of polariton discovery guided by symmetry considerations in polar crystals. After revisiting the underlying fundamental concepts, and displaying recent discoveries in the context of hyperbolic phonon polaritons (HPhPs) in natural crystals, we discuss the opportunities emerging from these concepts in the context of metamaterials, e.g., leveraging twist-optics and low-symmetry metacrystals.

\section*{Fundamental concepts for phonon polaritons in low-symmetry materials}

Phonon polaritons~\cite{caldwell2015low,Foteinopoulou2019PhononPolaritons} emerge in the infrared (IR) spectral range in polar dielectric crystals. In the absence of free charge carriers, the optical response in these materials is dominated by dynamic displacements of the ions in the crystal, i.e., optical phonons~\cite{bornhuang1954}. Polar lattices are characterized by partially charged ionic sites, such that optical phonon oscillations are associated with an induced polarization that is proportional to the partial charges or Born effective charges of the ions. For transverse optical (TO) phonons at the $\Gamma$-point of the Brillouin zone, this polarization can directly couple to light, making these modes IR-active. We emphasize that, owing to the small momentum of light in the IR spectral range, similar to Raman spectroscopy, it is typically sufficient to consider only $\Gamma$-point phonons. The corresponding longitudinal optical (LO) phonons, on the other hand, emerge at higher energies because of the associated macroscopic longitudinal polarization~\cite{bornhuang1954}. The magnitude of the TO-LO splitting at the $\Gamma$-point reflects the oscillator strength (or the magnitude of light-matter coupling strength) of the given phonon mode, and thus enters directly into the Lorentz-oscillator model that is typically employed to account for the optical phonon contribution to the linear permittivity:
\begin{equation}
    \varepsilon (\omega) = \varepsilon_\infty \left( 1 +  \frac{\omega_{LO}^2-\omega_{TO}^2}{\omega_{TO}^2-\omega^2-i\gamma_{TO}\omega} \right),
    \label{eq:phonon_oscillator}
\end{equation}
where $\omega_{TO(LO)}$ and $\gamma_{TO}$ are the frequency of the TO(LO) phonon and the phonon damping, respectively, and $\varepsilon_\infty$ the high-frequency electronic permittivity. For some crystals, the difference between LO and TO phonon damping needs to be accounted for using the four-parameter semi-quantum model~\cite{gervais1974anharmonicity} in order to satisfactorily reproduce the experimental data. An example of the resulting permittivity dispersion associated with Eq.~\ref{eq:phonon_oscillator} is shown in Fig.~\ref{fig:fig1}a. Here, the frequency dependence of $\varepsilon(\omega)$ reflects the non-instantaneous response of the material in the proximity of the phonon resonance. This model or its multi-mode equivalent~\cite{bornhuang1954} in the presence of multiple optical phonon resonances, fully describes the linear optical response of isotropic polar crystals, assuming the ion displacements to be small. As a direct consequence of Eq.~\ref{eq:phonon_oscillator}, the real part of the permittivity can turn negative between the TO and LO frequencies, constituting the bulk-polariton gap or Reststrahlen band (RB), within which light cannot propagate inside the crystal. Notably, with typical TO-LO splittings of 10-30~\% of the TO frequency, PhPs are naturally in the ultra-strong light-matter coupling limit (i.e. the coupling strength is comparable to the transition frequencies\cite{frisk2019ultrastrong}) for a large number of polar crystals, making them of significant relevance for polariton physics. The RB splits the dispersion relation that governs light propagation in the material, establishing a lower and an upper bulk PhP branch~\cite{bornhuang1954,henry1965raman}. Within the RB, light propagation is only associated with surface or interface modes that decay into the bulk~\cite{falge1973dispersion}, forming SPhPs. The dispersion of SPhPs at the interface between isotropic polar crystals and air is identical to that of surface plasmon polaritons~\cite{maier2007plasmonics}, with a wave number dependence on frequency described by
\begin{equation}
    k_\parallel(\omega) = \frac{\omega}{c_0}\sqrt{\frac{\varepsilon(\omega)}{1+\varepsilon(\omega)}},
    \label{eq:sphp_dispersion}
\end{equation}
where $k_\parallel$ is the complex-valued SPhP momentum parallel to the interface and $c_0$ is the speed of light in vacuum. Surface-bound solutions to Eq.~\ref{eq:sphp_dispersion} are found for $\Re(k_\parallel) > k_0$ where $k_0$ is the momentum of light in free space, implying that SPhPs are always evanescent and sub-diffractional. 

Notably, the above description based on Lorentz oscillators entering the permittivity constitutes the linear classical electromagnetic theory for polaritons~\cite{bornhuang1954}. We will follow this model in the following sections, introducing broken symmetries and phonon anisotropy in the permittivity tensor through the Lorentz oscillator model parameters. Higher-order nonlinear optical responses may become important when intense light fields are considered, especially considering that SPhPs tend to localize the optical fields in subwavelength regions. Higher-order susceptibility tensors are governed by higher-order group symmetries than the linear response~\cite{winta2018second}, yielding even more exotic responses in low-symmetry materials. Furthermore, the considerations above assume phonon and polariton wavelengths much larger than the relevant crystal lattice dimensions, implicitly assuming a local field approximation. Nonlocal phenomena~\cite{gubbin2020optical} manifest a wavevector-dependence of the permittivity, which becomes important for ultra-confined polaritons~\cite{correas2015nonlocal} and in atomic-scale heterostructures~\cite{caldwell2016atomic,ratchford2019CrystalHybrid}. 

Equations~\eqref{eq:phonon_oscillator}-\eqref{eq:sphp_dispersion} assume a crystal with isotropic optical response, as in the case of cubic crystals, such as cubic BN, exemplified in Fig.~\ref{fig:fig1}a. In this case, the permittivity can be treated as a scalar quantity. In the following, we are particularly interested in materials with a nontrivial permittivity tensor $\hat \varepsilon$, such that the polarization response $\vec P$ of a material to an external electric field vector $\vec E$ is written as $\vec P = \hat\varepsilon \vec E$~\cite{landau2013electrodynamics}, with
\begin{equation}
    \hat{\varepsilon} = 
    \begin{pmatrix} 
    \varepsilon_{xx} & \varepsilon_{xy} & \varepsilon_{xz} \\
    \varepsilon_{xy} & \varepsilon_{yy} & \varepsilon_{yz} \\
    \varepsilon_{xz} & \varepsilon_{yz} & \varepsilon_{zz}
    \end{pmatrix}.
    \label{eq:epstensor}
\end{equation}
The isotropic scenario is obtained as the special case of  $\hat\varepsilon$ being diagonal, with identical diagonal components $\varepsilon_{xx}=\varepsilon_{yy}=\varepsilon_{zz}$. This optical response stems from the transformation properties of the tensor, which in the isotropic scenario is invariant to rotation $R$ by any angle $\vartheta$ about an arbitrary axis $a$, $\hat\varepsilon_{iso} = R_a (-\vartheta) \hat\varepsilon_{iso} R_a(\vartheta)$~\cite{landau2013electrodynamics}. Notably, this implies that the optical response possesses a higher symmetry compared to the structural symmetry of the crystal, which is typically invariant only under discrete rotations about specific axes~\cite{ashcroft2022solid}. This higher symmetry is due to the local approximation of our model. In particular, for phonon resonances the polarization vectors for the vibrational normal modes are typically oriented along specific directions in the crystal structure. In this picture, the higher symmetry of the optical response is the consequence of a degeneracy of the underlying phonon resonators~\cite{tompkins2005handbook,weiglhofer2003introduction}.

When we reduce the crystal symmetry, as illustrated in the first row of Fig.~\ref{fig:fig1}, we generally also reduce the symmetry of the optical response. This is shown by a graphical representation of the permittivity tensor in the second row of Fig.~\ref{fig:fig1}, where the three arrows indicate the three diagonal entries of the permittivity tensor in the crystal frame, which may be rotated against an interface, e.g. the crystal surface (grey shaded area), see Fig.~\ref{fig:fig1}d. The reduced crystal symmetry results in the lifting of the degeneracy of the phonon resonators, illustrated by the color of the arrows, with examples of the permittivity along different crystal axes plotted in the bottom row of Fig.~\ref{fig:fig1}. For example, in hexagonal, trigonal and tetragonal crystals, material resonances are still degenerate in the plane normal to the $c$-axis, resulting in a uniaxial optical response where $\hat\varepsilon$ is still diagonal in the crystal frame. However, only two diagonal entries are identical and different from the third one, which constitutes the optical axis (OA) of the uniaxial material. This degeneracy lifting results in a dependence of the optical response of an interface or of a thin film on its orientation with respect to the OA (see Fig~\ref{fig:fig1}b-d). Note that the permittivity tensor for uniaxial materials is still invariant to continuous rotations around the OA, despite the fact that the underlying hexagonal, trigonal, or tetragonal crystal lattices obey only discrete 6-,3-, or 4-fold rotations, respectively. For instance, with the c-axis perpendicular to the interface (Fig.~\ref{fig:fig1}b) the optical response is only in-plane isotropic, while more complex responses can be expected for in-plane (Fig.~\ref{fig:fig1}c) or even canted OAs (Fig.~\ref{fig:fig1}d). 

A further reduction in crystal symmetry leads to non-degenerate phonon resonances along all three crystal axes in orthorhombic crystals, producing a biaxial optical response (Fig.~\ref{fig:fig1}e). The permittivity tensor is still diagonal in the crystal frame, despite exhibiting different entries in all three components $\varepsilon_{xx} \neq \varepsilon_{yy} \neq \varepsilon_{zz}$, such that $\hat \varepsilon$ is no longer invariant under any non-trivial rotation, but it retains mirror symmetries about the crystal planes. Finally, in monoclinic and triclinic crystals, the transition dipole moments of the phonon modes are no longer parallel to any of the crystal axes~\cite{schubert2016galliumoxide,schubert2017CdWO4}, such that there is no reference frame in which the full complex permittivity tensor can be cast in a diagonal form within the monoclinic plane (Fig.~\ref{fig:fig1}f), or overall for triclinic crystals (Fig.~\ref{fig:fig1}g). Mirror symmetries are broken in the monoclinic plane (Fig.~\ref{fig:fig1}f), vanishing entirely for triclinic crystals (Fig.~\ref{fig:fig1}g). 

The most important implication of strong and directional phonon resonances in anisotropic polar crystals for infrared nanophotonics is the emergence of natural hyperbolicity~\cite{dai2014tunable,caldwell2014sub,ma2018plane,passler2022hyperbolic}. When different principal components of the permittivity tensor have opposite sign, as naturally occurring near directional phonon resonances, these materials support hyperbolic polariton propagation in the bulk. Hyperbolic polaritons are characterized by an open topology in the corresponding isofrequency contours (IFCs) providing extreme light confinement, high density of photonic states and highly directional propagation~\cite{dai2014tunable,caldwell2014sub,liu2007far}. In the following sections, we review the pronounced effects of crystalline and optical symmetry breaking on hyperbolic polaritons, discussing the underlying principles and outlining the emerging opportunities for polaritonics.

\section*{Hexagonal crystal system with uniaxial optical response}
\label{sec:uniaxial}

\subsection*{Out-of-plane and in-plane uniaxial crystals}
\label{sec:outofplaneuniaxial}

\begin{figure}
    \centering
\includegraphics[width=1\textwidth]
{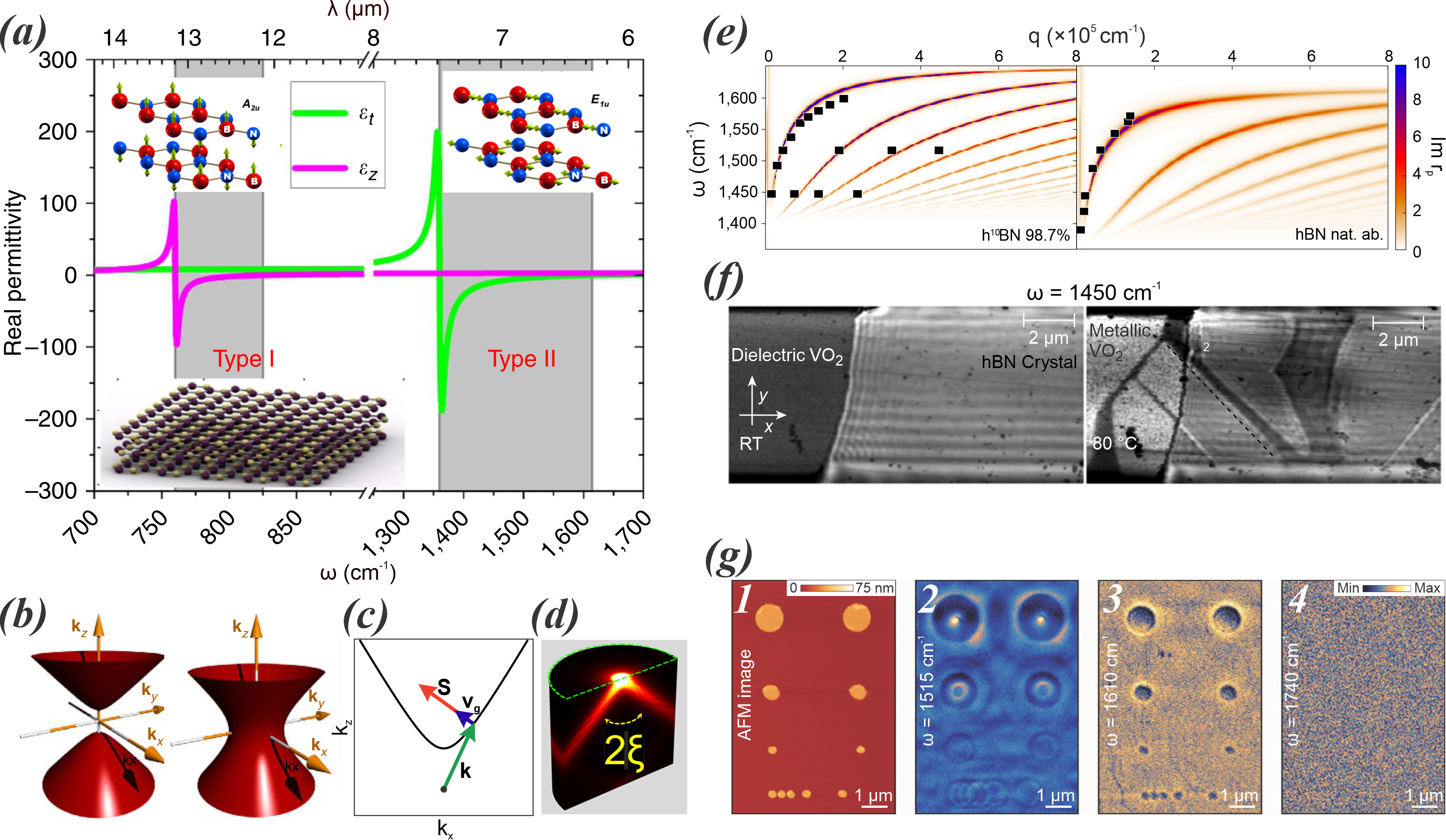}
    \caption{\textbf{Out-of-plane hyperbolicity in uniaxial crystals: volume-hyperbolic phonon polaritons in hBN}. \textbf{(a)} In-plane (green, $\varepsilon_x=\varepsilon_y=\varepsilon_\perp$) and out-of-plane (magenta, $\varepsilon_\parallel$) components of the dielectric permittivity tensor of hBN (inset), leading to type I ($\varepsilon_\parallel<0$,$\varepsilon_\perp>0$) and type II ($\varepsilon_\parallel>0$,$\varepsilon_\perp<0$) hyperbolicity. The insets show the atomic displacements associated with the out-of-plane ($A_{2u}$) and in-plane ($E_{1u}$) IR-active phonon resonances in the lower and upper RBs, respectively~\cite{gil2020boron}. \textbf{(b)} Corresponding isofrequency surfaces for the (left) type I and (right) type II hyperbolic regimes~\cite{caldwell2014sub}. Panel \textbf{(c)} illustrates the non-collinearity between the $\mathbf{k}$-vector and the group velocity $v_g$ typical of hyperbolic polaritons. \textbf{(d)} Ray-like directional propagation of HPhPs from a point source~\cite{li2017optical}. The propagation angle $\xi$ is determined by the hyperbola asymptote, see Eq.~\ref{eq:propag_angle}. \textbf{(e)} Dispersion curves in the type II hyperbolic RB for: $98.7\%$ $^{10}B$ (left panel), and naturally abundant (right panel) hBN~\cite{giles2018ultralow}. The calculations (color plots), overlaid by experimental data (black squares), are extracted from the dielectric function fit to the experimental reflectance spectra. \textbf{(f)} Experimental s-SNOM images of an hBN flake on top of a $VO_2$ substrate. Upon temperature increase (room temperature and 80$^\circ$, left and right, respectively), HPhPs are refracted at the VO$_2$ phase boundaries~\cite{folland2018reconfigurable}. \textbf{(g)} Super-resolution imaging of gold nanodisks through an hBN slab~\cite{dai2015subdiffractional}. The s-SNOM images shown in panel 2 and 3 are taken at frequencies $\omega = 1515~cm^{-1}$ and $\omega = 1610~cm^{-1}$, respectively, both within the upper RB of hBN, while no imaging contrast is observed outside the RB at $\omega = 1740~cm^{-1}$, shown in Panel 4.}
    \label{fig:fig2}
\end{figure}

Uniaxial polar crystals include a plethora of layered two-dimensional materials belonging to the hexagonal crystal family~\cite{born2013principles}. Among them, hexagonal boron nitride (hBN) is one of the most extensively employed and investigated materials, owing to its large band gap, which makes it an ideal encapsulant for graphene and other VdW materials~\cite{caldwell2019photonics}. Its crystalline structure, consisting of stacked VdW-bound layers of boron and nitrogen arranged into hexagonal lattices with strong in-plane covalent bonds, features two IR-active phonon resonances: a doubly degenerate in-plane vibrational mode ($E_{1u}$) at high frequency, and a low-frequency out-of-plane mode ($A_{2u}$) dominated by interlayer atomic vibrations (Fig.~\ref{fig:fig2}a) along the c-axis (perpendicular to the VdW-bound layers). 

This asymmetry in the crystal lattice dynamics is reflected in the optical response, resulting in drastically different dielectric permittivity components along the two major polarizability axes, see Fig.~\ref{fig:fig2}b. Light polarized orthogonal (parallel) to the c-axis\cite{gil2020boron} exclusively couples to the $E_{1u}$  ($A_{2u}$) modes, resulting in birefringence of hBN near these phonon resonances. Thus, hBN supports bulk propagation of ordinary and extraordinary waves~\cite{landau2013electrodynamics}. The ordinary waves exhibit spherical IFCs akin to those in the bulk of isotropic media, where the momentum and wavelength are independent of the propagation direction. The dispersion relation for extraordinary waves, on the other hand, reads~\cite{su2021surface,caldwell2014sub}:
\begin{equation}
    \frac{k_x^2+k_y^2}{\varepsilon_\parallel} + \frac{k_z^2}{\varepsilon_\perp} = k_0^2,
    \label{eq:eq_disp_bulk_hBN}
\end{equation}
where $k_{x,y,z}$ are the wave momentum components in Cartesian coordinates $x,y,z$, and $\varepsilon_\parallel =\varepsilon_z$ and $\varepsilon_\perp = \varepsilon_{x,y}$ are the permittivity components parallel and perpendicular to the OA, respectively. 

Spectrally away from directional resonances, the out-of-plane and the in-plane dielectric tensor components ($\varepsilon_\parallel$ and $\varepsilon_\perp$, respectively) are real-valued (in the absence of material loss) and have different magnitudes, hence the IFCs become ellipsoidal, reflecting the anisotropy in the optical response~\cite{landau2013electrodynamics}. Interestingly, the birefringence of hBN becomes extreme within its RBs, where the two permittivity components acquire opposite signs (Fig.~\ref{fig:fig2}b): $Re(\varepsilon_\perp) \cdot Re(\varepsilon_\parallel) < 0 $. Inside these frequency regions, the dispersion relation for the extraordinary wave in Eq.~\ref{eq:eq_disp_bulk_hBN} in the lower (upper) RB of bulk hBN consists of a two-sheet (one-sheet) open hyperboloid in $\mathbf{k}$-space (Fig. \ref{fig:fig2}c), leading to the formation of type I (type II) HPhPs~\cite{su2021surface,caldwell2014sub}. These hyperbolic waves showcase a range of exotic features that make them highly appealing for nanophotonic applications.

Firstly, hyperbolic IFCs have an open topology, hence they can in principle access very large wave vectors, associated with highly confined fields, extreme light-matter interactions, as well as ultra-slow light propagation~\cite{basov2021polaritonpanorama}. The highly sub-diffractive nature of hyperbolic polaritons is only limited by losses, which are enhanced for highly confined fields and at some point limit the maximum wave number accessible in realistic materials, and by nonlocalities~\cite{correas2015nonlocal}. Further, the power flow for hyperbolic waves is highly collimated. This can be recognized by noticing that, in the $\mathbf{k}$-space representation considered in Figure \ref{fig:fig2}c, the power flow, associated with the Poynting vector $\Vec{S}=\Vec{E} \times \Vec{H}$, is orthogonal to the isofrequency surfaces. For isotropic media this implies that phase propagation (real part of $\mathbf{k}$-vector) and power flow are parallel as expected. For hyperbolic dispersion, however, the phase velocity vector is parallel to the Poynting vector only at the vertices (in type I hyperbolic bands) or along the waist (in type II hyperbolic bands) of the dispersion hyperboloids. The two vectors actually become orthogonal at the asymptotes~\cite{caldwell2014sub}. In the asymptotic regime for large k-vectors, a continuum of momentum vectors share the same power flow direction, giving rise to a highly directional power flow. Thus, not only do these waves propagate into the bulk within the otherwise prohibited region (RB)\cite{jacob2014hyperbolic}, but they do so in a directional ray-like fashion (see Fig.~\ref{fig:fig2}d). Their energy flow is mostly directed at an angle~\cite{dai2014tunable,caldwell2014sub,li2015hyperbolic}:
\begin{equation}
    \xi(\omega) = \pi/2 - \arctan{(\sqrt{\varepsilon_\parallel(\omega)}/i\sqrt{\varepsilon_\perp(\omega)})},
    \label{eq:propag_angle}
\end{equation}
defined as the angle between the Poynting vector and the OA as a function of the principal components of the dielectric tensor (see Fig.~\ref{fig:fig2}d~\cite{li2017optical}), when approximating Eq.~(\ref{eq:eq_disp_bulk_hBN}) by a cone\cite{caldwell2014sub, li2015hyperbolic}. 

These remarkable features which extend across the entire RBs of phonon polariton materials make hyperbolic responses particularly appealing for nanophotonics, due to the extreme light-matter interactions and deeply sub-diffractive propagation they enable. Notably, the asymptotic angle Eq.~\ref{eq:propag_angle} strongly depends on frequency within each RB\cite{li2015hyperbolic} and converges towards 0 as the frequency approaches the TO (LO) phonon for type I (type II) hyperbolic regimes, giving rise to canalized rays propagating along the OA direction. This so-called canalization regime~\cite{gomez2015hyperbolic,belov2005canalization} of highly directional and diffractionless propagation corresponds to the $\varepsilon$-near-zero canalization regime discussed in the context of metamaterials~\cite{correas2017plasmon}. 

Owing to its weak interlayer VdW forces, hBN flakes can be easily exfoliated and transferred onto different substrates. When considering an ultra-thin ($d << \lambda_0$, where $\lambda_0=2\pi c_0/\omega$) hBN slab sandwiched between an optically isotropic substrate, such as silicon, and air, the hyperbolic polariton propagates in a ray-like fashion along the direction defined by Eq.~\ref{eq:propag_angle} until it reaches the opposite interface and bounces back like in a Fabry-Perot-cavity. The in-plane dispersion relation for such volume-confined hyperbolic polaritons (v-HPhPs) in ultra-thin sheets shows multiple thickness-dependent branches~\cite{dai2015subdiffractional,dai2014tunable}. Under the large momentum approximation $|\mathbf{k}|\gg k_0$, the resulting modal dispersion can be written as \cite{AlvarezPerez19}
\begin{equation}
    k_\parallel(\omega) = -\frac{\psi(\omega)}{h}\left\lbrace\text{arctan}\left[\frac{\varepsilon_1}{\varepsilon_\perp(\omega) \psi(\omega)}\right]+\text{arctan}\left[\frac{\varepsilon_2}{\varepsilon_\perp(\omega) \psi(\omega)}\right]+\pi l\right\rbrace,
    \label{eq:disp_ultrathin_hBN}
\end{equation}
where $l = 0,1,2, ...$ is the mode index, $\psi(\omega) = \sqrt{\varepsilon_\parallel(\omega)}/i \sqrt{\varepsilon_\perp(\omega)}$, and $h$ is the hBN flake thickness, see for instance Fig.~\ref{fig:fig2}e. Notably, Eq.~\ref{eq:disp_ultrathin_hBN} predicts a negative slope of the dispersion in type-I hyperbolic ranges, leading to a negative phase velocity~\cite{yoxall2015direct}.

%Whilst in the lower RB the dispersion branch of the fundamental mode $M_0$ tends asymptotically to the TO phonon frequency, in the upper hyperbolic region it approaches the LO phonon. This feature indicates at which end of the RB the HPhP is most confined in the two cases, namely at the TO (LO) resonance for the lower (upper) branch. Contrary to previous understanding~\cite{li2015hyperbolic,caldwell2014sub,dai2015subdiffractional} that had attributed the different trends to opposite signs of the group velocity $v_g = \frac{\partial \omega}{\partial k}$, this behaviour has later been explained by the negative phase velocity $v_p$ in the type I hyperbolic region.
%This exotic phenomenon arises as a consequence of the previously mentioned non-collinearity between the power flow direction and the $\mathbf{k}$-vector, and hence between $v_g$ and $v_p$. In this perspective, the negative sign of the dispersion branches results from the negative $k_x$ with respect to the Poynting vector. Although the projection of $k_x$ decreases in amplitude upon rotation of $\Vec{S}$ away from the c-axis with increasing incident frequencies (as shown in Fig.~\ref{fig:fig2}f), its negative sign ensures a positive dispersion gradient (group velocity), at the same time leading to a negative phase velocity 
%$v_p = \frac{\omega}{k}$\cite{yoxall2015direct}. 

The group velocity of HPhPs in hBN can approach very small values  $\approx 0.002$c and $\approx 0.027$c in type I and type II hyperbolic regions, respectively~\cite{yoxall2015direct}. This feature arises from the large wave vectors and associated small gradients accessed by the hyperbolic dispersion, see Fig.~\ref{fig:fig2}e. However, such low group velocities can only be discerned due to the high quality factor arising from the very long lifetimes of HPhPs in hBN, reaching several ps~\cite{low2017polaritons,caldwell2015low}. Following Eq.~\ref{eq:disp_ultrathin_hBN}, the mode confinement as well as the slope of the dispersion lines, and hence the group velocity, can be controlled by varying the thickness of the layers~\cite{dai2014tunable,li2021direct}. However, the layer thickness does not change the modal lifetime $\tau_p$, such that reducing  the group velocity also leads to a reduction of the propagation length $L_p = v_g \tau_p$.

The long lifetime of HPhPs is due to the low damping rate $\gamma_{TO}$ of TO phonons in polar crystals. In fact, in contrast to metals, where the damping rate is dictated by electronic losses resulting in typical lifetimes of only tens of fs, in polar crystals this factor is dominated by the phonon scattering rate, which represents the upper limit for the polariton lifetime~\cite{low2017polaritons,khurgin2018phononplasmon}. Polariton lifetimes of 1.8 ps and 0.8 ps were measured in a hBN thin slab on a SiO$_2$ substrate for the lower and upper RBs, respectively. Furthermore, a reduction of the scattering rates and the consequent increase in hBN HPhP lifetimes and propagation lengths was demonstrated via isotopic enrichment~\cite{giles2018ultralow}. The resulting increase in the HPhP quality factors in isotopically pure hBN crystals, compared to naturally abundant hBN, is displayed in Fig.~\ref{fig:fig2}e, accompanied by small shifts of the phonon frequencies and, as a consequence, the RB.

Since the discovery of HPhPs in hBN, new heterostructures have been developed to control their properties. For instance, flakes of isotopically enriched hBN have been transferred onto substrates of VO$_2$, a material that undergoes a dielectric-to-metal transition as its temperature is changed from $60^\circ C$ to $80^\circ C$\cite{folland2018reconfigurable}. Fig.~\ref{fig:fig2}f depicts the propagation of v-HPhPs inside the hBN slab placed over the dielectric (dark gray) and metallic (light gray) domains of VO$_2$. Heating up the substrate provides the opportunity to "switch on" the metallic domains, resulting in shorter HPhP wavelengths in the brighter metallic domains. Similarly, the confinement of HPhPs can be tuned by placing a graphene monolayer onto an hBN slab~\cite{dai2015graphene}, resulting in hybrid plasmon-phonon polaritons. Even more appealing, applying a gate voltage to the graphene monolayer enables active tuning of the hybrid polariton wavelength~\cite{caldwell2019photonics}.

In the past years, v-HPhPs in out-of-plane uniaxial crystals have found fruitful applications in nano-optics thanks to their high directionality and extreme sub-wavelength confinement. For instance, the use of these polaritons for super-resolution imaging in the Type II hyperbolic band of hBN has been demonstrated by imaging an array of Au nanodisks through an hBN slab~\cite{li2015hyperbolic,dai2015subdiffractional}, see for instance Fig.~\ref{fig:fig2}g. Importantly, the sub-diffractional gold nanodisks underneath the hBN layer can be resolved across the entire RBs (Fig.~\ref{fig:fig2}, panel 1-3), as the HPhPs carry high momentum components necessary to experimentally resolve the sub-diffractional images (Fig.~\ref{fig:fig2}k, panel 4)~\cite{li2015hyperbolic,dai2015subdiffractional}. Effectively, the hyperbolic volume-polariton rays act as a magnifying lens as shown in Fig.~\ref{fig:fig2}g for these Au disks, whose size $d$ can be reconstructed by measuring the size $D$ of the enlarged image via the formula $D=d+2h\tan{\xi}$, where $h$ is the thickness of the hBN flake and $\xi(\omega)$ is the dispersive propagation angle of the v-HPhP in Eq.~\ref{eq:propag_angle}~\cite{li2015hyperbolic}. The concentric rings visible on the top surface of hBN (displayed in panel 2) arise from the multiple reflections that occur at the slab's interfaces, and are a result of the superposition of several discrete modes simultaneously launched in $\mathbf{k}$-space.
Another interesting application of v-HPhPs is molecular sensing~\cite{autore2018boron,bylinkin2021real}. Many molecular vibrations have their resonances in the same spectral region as phonon polaritons in VdW crystals. Notably, in contrast to plasmonic sensing approaches~\cite{stewart2008nanostructured}, the high quality factors of HPhPs result in sensing schemes that rely on strong coupling~\cite{autore2018boron} rather than enhancement of vibrational absorption.   

%the presence of molecules underneath a hBN slab gives rise to a hybridized state, whose dispersion is characterized by avoided crossing regions. The lines of reduced electric field amplitude in Fig.~\ref{fig:fig2}l (right panel) result from vibrational strong coupling with 4,4$^\prime$-bis(N-carbazolyl)-1,1$^\prime$-biphenyl (CBP) molecules, and correspond to the resonance frequencies (that is, the fingerprints) of the molecular vibrations. 

Despite supporting ultra-confined HPhPs with high out-of-plane directionality and other intriguing features, hBN - alongside many VdW materials - inherently only provides out-of-plane hyperbolicity. Its layered VdW-bound structure limits crystal cleaving to the VdW-bound plane, and thus HPhP propagation is always isotropic in the interface plane. More traditional uniaxial crystals such as $\alpha$-quartz\cite{falge1973dispersion,winta2019low} or calcite\cite{ma2021ghost,ni2023observation} offer more options since the crystals can be cut in arbitrary orientation. Whilst relatively unexplored, recent works showed in-plane hyperbolicity with highly directional propagation for $y$-cut quartz~\cite{passler2022hyperbolic}, as originally introduced in the context of metasurfaces\cite{high2015visible,gomez2015hyperbolic}. Moreover, entirely new forms of polaritons have been observed recently using off-cut uniaxial crystals~\cite{ma2021ghost,ni2023observation}, as discussed in detail in the following section.

\subsection*{Off-cut uniaxial crystals}
\label{sec:offcut}

\begin{figure}
    \centering
\includegraphics[width=\textwidth]{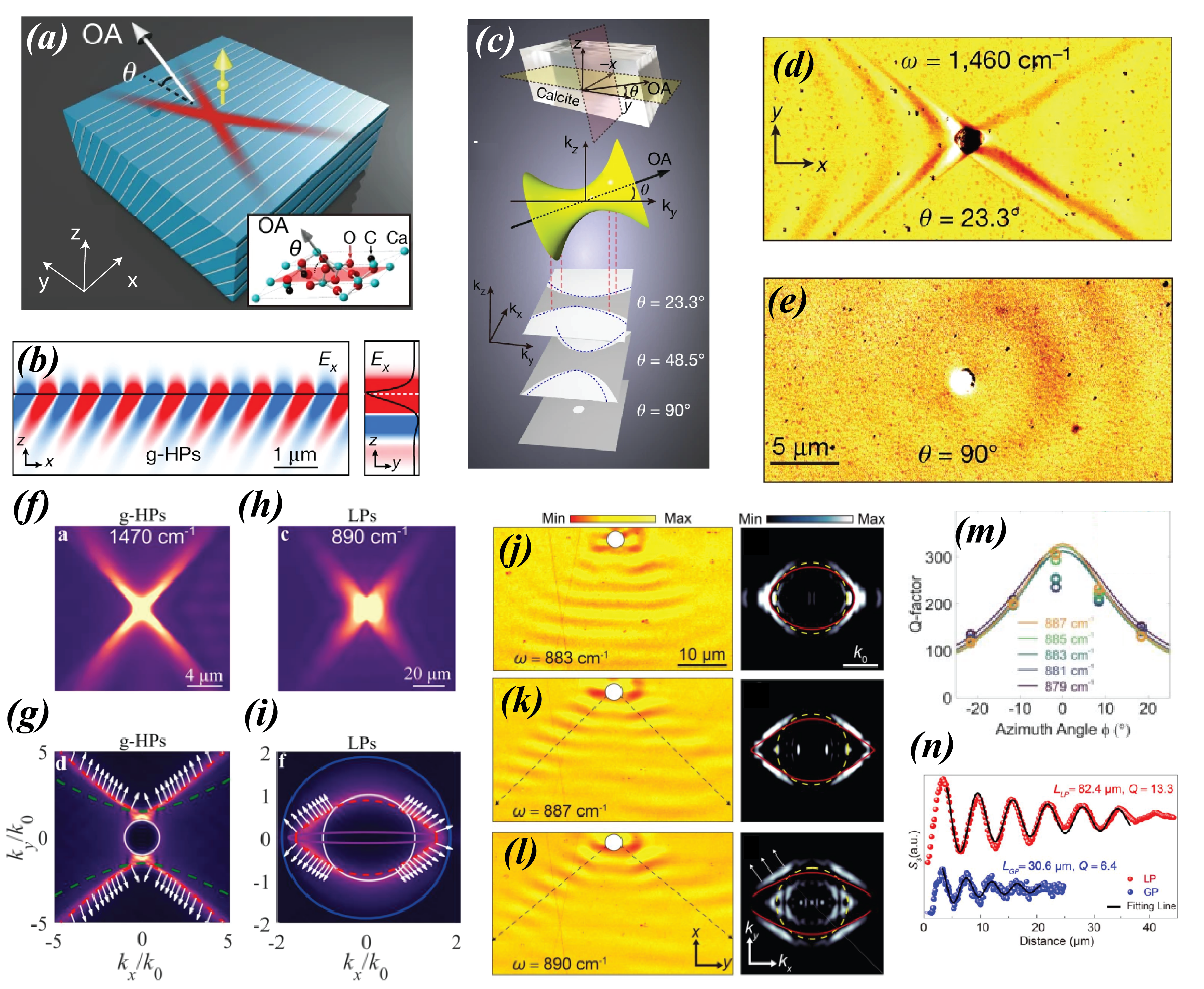}
    \caption{\textbf{Surface anisotropy in off-cut crystals: ghost polaritons and leaky polaritons}. \textbf{(a)} Illustration of HPhPs launched by a point-dipole placed near the surface of an off-cut uniaxial crystal with its OA slanted relative to the surface in its type-II hyperbolic frequency region (upper RB). \textbf{(b)} Theoretically predicted ghost polaritons in off-cut calcite, whose out-of-plane wavevector possesses both a real and an imaginary component, resulting in combined oscillatory and decaying behavior into the anisotropic material. \textbf{(c)} OA orientation (top) relative to the surface cut and (bottom) in-plane polariton dispersion (blue dashed lines) for experimentally accessible cut angles $\theta=23.3^\circ$, $48.5^\circ$ and $90^\circ$. The latter is only out-of-plane uniaxial, and hence exhibits an elliptic dispersion contour in the $k_x-k_y$ plane. \textbf{(d)} Experimental s-SNOM measurement of ghost polaritons in 23.3$^\circ$ offcut calcite, launched by a gold nano-disk at frequency $\omega=1,460$ cm$^{-1}$, and \textbf{(e)} corresponding measurement for the $\theta=90^\circ$ sample, which does not support ghost polaritons 
    \cite{ma2021ghost}. \textbf{(f,h)} Near-field distribution of \textbf{(f)} ghost polaritons and \textbf{(h)} leaky polaritons from point-dipole simulations, showing the in-plane directionality of polaritons in real space. \textbf{(g)} Fourier spectra of the hyperbolic ghost polariton shown in \textbf{(f)}, overlaid with its analytic hyperbolic contour. \textbf{(i)} Fourier spectra of the leaky polariton shown in \textbf{(h)}, overlaid with its analytic lenticular contour. White-colored arrows in \textbf{(g)} and \textbf{(i)} denote the Poynting vectors of the respective polaritons, highlighting their strong directionality. \textbf{(j-l)} Near-field imaging of leaky polaritons at various frequencies (left panels) and their corresponding FTs (right panels). \textbf{(m)} The quality factor of leaky polaritons at multiple frequencies extracted from Otto-type prism coupling measurement. \textbf{(n)} Comparison between the damping of ghost polaritons (red-colored) and leaky polaritons (blue colored) measured by s-SNOM~\cite{ni2023observation}.}
    \label{fig:fig4}
\end{figure}

Further lowering the symmetry, we can study the polaritonic response at the surface of uniaxial materials when these are cut at a finite angle $\theta$ relative to their OA. One material that can be cut in this way is calcite. Starting from the $x'-y'-z'$ reference frame in which the material is described by a diagonal uniaxial dielectric tensor $\hat\varepsilon' = diag{[\varepsilon_\perp,\varepsilon_\parallel,\varepsilon_\perp]}$, the sample can be cut in the $y-z$ plane at an angle $\theta$ with respect to the $y$ axis, as shown in Fig.~\ref{fig:fig4}a. By rewriting the dielectric tensor in the $x-y-z$ frame as $\hat\varepsilon = \hat{R}_x(-\theta)\hat\varepsilon'\hat{R}_x(\theta)$, where the $z$ axis is normal to the surface and $\hat{R}_x(\theta)$ is a rotation matrix, an off-diagonal component arises that couples in-plane and out-of-plane polarizations\cite{ma2021ghost}. Importantly, this coupling is not an effect of the broken crystal symmetry, but simply of its orientation with respect to the interface.

The resulting dispersion relation for extraordinary waves in this system reads
\begin{equation}
    \frac{(k_y\cos\theta-iq_{ze}\sin\theta)^2}{\varepsilon_\perp} + \frac{k_x^2+(k_y\sin\theta+iq_{ze}\cos\theta)^2}{\varepsilon_\parallel}=k_0^2,
\end{equation}
where $q_{ze}$ is the out-of-plane decay constant of extraordinary waves. Different solutions for polaritons bound to the interface exist, and possess fundamentally different character, depending on the quantities $\Xi = \varepsilon_\parallel\sin^2\theta+\varepsilon_\perp\cos^2\theta$, and 
\begin{equation}
    \Delta = \frac{k_y ^2}{\varepsilon_{\parallel}\sin^2\theta+\varepsilon_\perp\cos^2\theta}+\frac{k_x^2}{\varepsilon_\parallel}-k_0^2.
\end{equation}
For $\Xi < 0$, and $\Delta > 0$, a new form of polariton emerges. If $0<\theta<90^\circ$, the decay constant $q_{ze}$ becomes complex, implying simultaneous oscillation and decay away from the interface, in sharp contrast with Dyakonov surface waves\cite{dyakonov1988new} or conventional polaritons, whose evanescent character is purely exponential. These exotic surface waves, called ghost polaritons, were first theorized in 2019\cite{narimanov2018dyakonov,narimanov2019ghost}, and owe their name to an analogy with ghost orbit bifurcation in semiclassical quantization~\cite{kus1993prebifurcation}. They were observed for the first time in calcite in 2021~\cite{ma2021ghost}. Figure~\ref{fig:fig4}b shows the peculiar field profile of a ghost-HPhP, with its characteristic tilted phase front, demonstrating the damped oscillatory behavior of the ghost waves. Note that despite their tilted wavefronts, the Poynting vector and power flow of ghost polaritons is parallel to the interface~\cite{ma2021ghost}.

Figure~\ref{fig:fig4}c shows a sketch of the off-cut configuration in (top) real and (middle) Fourier space, with the definition of the tilt angle $\theta$ of the OA in the $y-z$ plane. The bottom three panels show in-plane projections of the polariton dispersion for $\theta = 23.3^\circ$, $48.5^\circ$ and $90^\circ$, as the ghost polariton dispersion undergoes a transition from hyperbolic to elliptical isofrequency contours. Quite remarkably, the polariton propagation for a fixed frequency can experience a topological transition from hyperbolic to elliptical IFCs, as a function of the angle between the interface and the OA. Figure~\ref{fig:fig4}d shows a near-field measurement of the ghost polariton excited by a gold disk in off-cut calcite with $\theta=23.3^\circ$, while panel (e) shows the field map for $\theta = 90^\circ$, where no ghost polariton exists and the propagation is effectively isotropic. Remarkably, the two images refer to the same frequency and the same material system, with the only difference being the angle $\theta$.

Another exciting class of phonon polaritons recently observed in off-cut calcite are leaky polaritons. Leaky polaritons hybridize with the refractive bulk modes of calcite, while being confined to the interface with air. Thus, while being mostly confined to the interface, they leak a portion of their power into the bulk of the anisotropic crystal, retaining high directionality, long propagation range and sub-diffractional features at the interface~\cite{ni2023observation}. As opposed to hyperbolic polaritons (Fig.~\ref{fig:fig4}f-g), which derive their high directionality and field confinement from the asymptotic large-momentum behaviour of their hyperbolic contours, the high directionality of leaky polaritons stems from the lenticular shape of their dispersion contour, which hosts a large number of collinear states that mimic hyperbolic propagation (Fig.~\ref{fig:fig4}h-i). Remarkably, leaky polaritons preserve their high directionality despite their proximity to the light cone of free space. In the case of calcite, they occur within the light cone of the ordinary waves hosted in the bulk of the polaritonic medium, but outside of the light cone of its extraordinary waves. Although both ghost and leaky polaritons are supported by calcite (albeit in different frequency regimes), and featuring a complex out-of-plane wavevector, the nature of ghost and leaky polaritons is very different. Ghost polaritons support a complex-valued out-of-plane momentum component $q_{ze}$ associated with tilted wave fronts, but their in-plane momentum is real-valued in the absence of material loss, such that the power flow is nevertheless parallel to the interface. 
By contrast, the wavevector of leaky polaritons is instead complex-valued also in the interface plane, due to the radiation leakage, resulting in both wavefront and power flow being slanted away from the interface.

Near-field measurements of leaky phonon polaritons were recently performed in 23.3$^\circ$ off-cut calcite, shown in Fig.~\ref{fig:fig4}j-l (left), along with their respective Fourier transforms (FTs, right), for three frequencies within or near the lower RB, showing the increased directionality enabled by their lenticular dispersion. The leaky nature of these polaritonic modes weakens their confinement, giving them longer propagation distances and lower susceptibility to loss, despite their radiation leakage into the bulk of calcite. This endows them with high quality factors, longer propagation lengths and lifetimes, even compared to conventional phonon polaritons. For instance, measured values for the quality factors relating to the lifetime of leaky phonon polaritons can be as high as $Q\approx 300$, as shown for different frequencies in Fig.~\ref{fig:fig4}m as a function of the azimuthal angle $\phi$, based on prism excitation via Otto configuration. Leaky polaritons also exhibit longer propagation lengths compared to ghost polaritons, as demonstrated in Fig.~\ref{fig:fig4}n, where near-field measurements of the amplitudes of a leaky polariton (in red), and a ghost polariton (in blue) are fitted to a decaying sinusoid, extracting the respective relative propagation lengths of $\approx 13.3$ and $6.4$. Note that leaky polaritons can exist not only in off-cut bianisotropic materials, but also in a wide range of anisotropic materials, even when the OA is parallel to the interface. For instance, recently performed Otto-type prism coupling measurements on y-cut $\alpha$-quartz, with OA oriented along the interface, revealed leaky polaritons in its type-I hyperbolic RB~\cite{ni2023observation}.

\section*{Orthorhombic crystal system with biaxial optical response}
\label{sec:biaxial}

Orthorhombic crystal structures are characterized by three lattice vectors $\vec{a}$, $\vec{b}$, and $\vec{c}$ which are perpendicular but have unequal lengths ($a \neq b \neq c$). As a result, the dipole moments and resulting lattice vibrations differ along each of these directions, and the degeneracy of the in-plane phonons observed in uniaxial materials is no longer present in orthorhombic crystals, thereby introducing an additional level of asymmetry. Consequently, there are two different OAs, and the optical response is thus referred to as biaxial. The dielectric tensor of a biaxial system can be diagonalized in an orthonormal basis that remains unchanged with frequency, due to the perpendicular arrangement of the lattice vectors in orthorhombic structures across all frequencies. The three components of the permittivity tensor along the principal directions differ: $\varepsilon_{xx} \neq \varepsilon_{yy} \neq \varepsilon_{zz}$. By contrast, and unlike crystals with lower symmetry, the off-diagonal components vanish: $\varepsilon_{xy}=\varepsilon_{yz}=\varepsilon_{xz}=0$. As a consequence of biaxial anisotropy, both bulk modes of the electromagnetic field supported by such media are extraordinary waves~\cite{AlvarezPerez19}, meaning that, in contrast with uniaxial media, neither of their dispersion curves follows a relation of the form $|k|=n\omega/c$, akin to that of isotropic media (here $n$ is a scalar refractive index).

\begin{figure}
    \centering
    \includegraphics[width=0.9\textwidth]{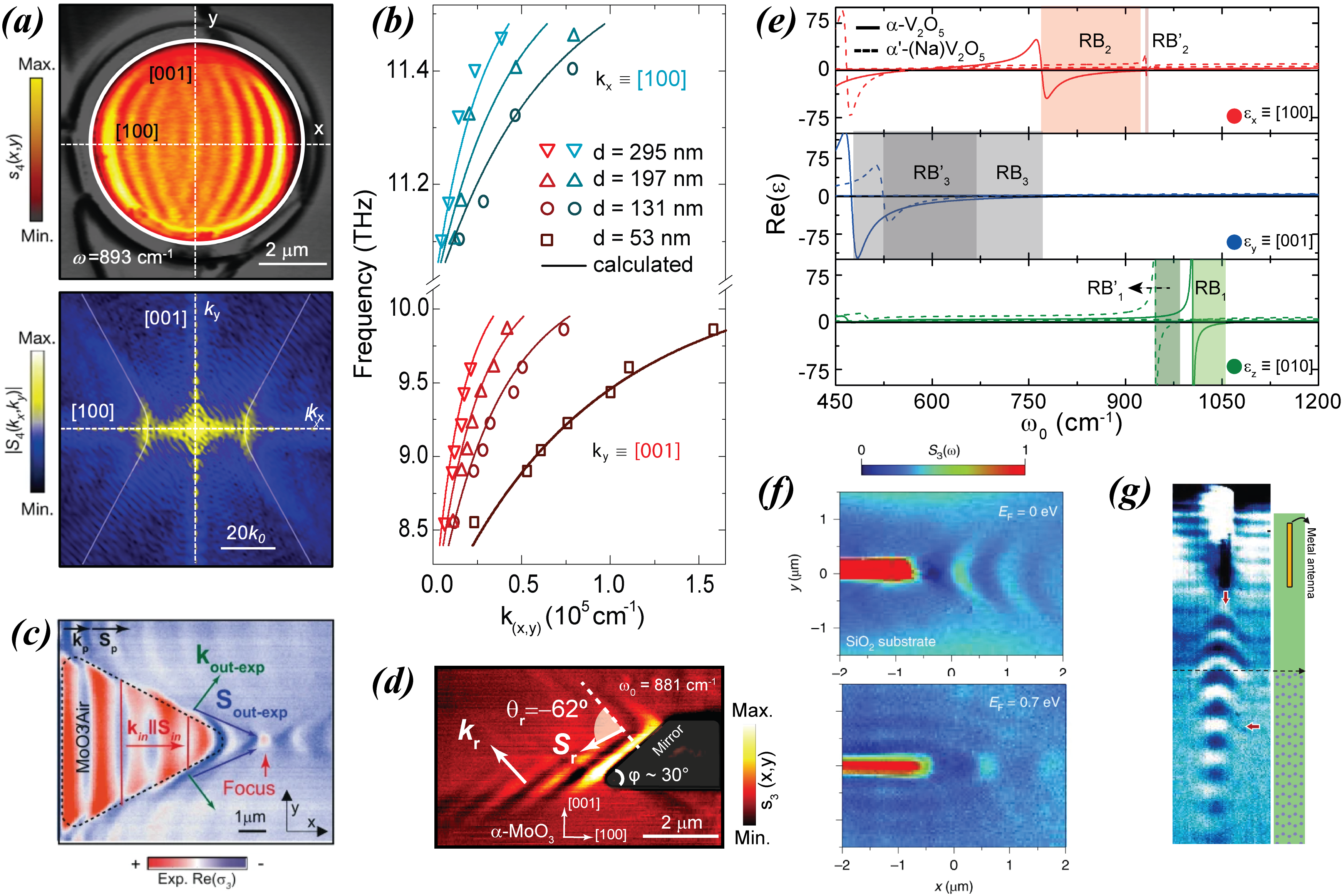}
    \caption{\textbf{Phonon polaritons in orthorhombic crystals with biaxial optical response}. \textbf{(a, top panel)} Near-field amplitude image $s_4$ of a 144-nm-thick $\alpha$-MoO$_3$ disk at $\omega_0= 893$ cm\textsuperscript{-1}. Dashed white lines indicate the [100] and [001] crystal directions. Scale bar: 2 $\mu$m. \textbf{(a, bottom panel)} Absolute value of the FT $|s_4(k_x,k_y)|$ of the near-field image in the top panel, revealing the IFC. Solid lines show the fitted PhP IFC (note that they correspond to $2k$ as they are tip-launched waves). Scale bar: $20 k_0$, with $k_0$ the momentum of light in free space~\cite{ma2018plane}. \textbf{(b)} THz polariton dispersion in $\alpha$-MoO$_3$ flakes of different thicknesses $d$. Data points and analytical dispersion are color-coded in shades of blue (for PhP\textsubscript{[100]}), and red (for PhP\textsubscript{[001]})~\cite{AlvarezPerez19}. \textbf{(c)} Experimental near-field image of a refractive planar hyperlens (black dashed line) for polaritons in a 170-nm-thick $\alpha$-MoO$_3$ slab, at $\lambda_0 = 11.16$ $\mu$m. HPhPs converge upon refraction at the triangular boundary. Compared to non-refracted polaritons, indicated by $\textbf{S}_p$ and $\textbf{k}_p$ (black arrows) the refracted polaritons, \textbf{S}\textsubscript{out-exp} (blue arrows), propagate nearly parallel to the boundary~\cite{Duan21}. \textbf{(d)} Experimental s-SNOM amplitude image $s_3(x,y)$ of HPhPs negatively reflecting on mirrors tilted at an angle of $\varphi = 38^\circ$ fabricated on $\alpha$-MoO$_3$. \textbf{k}$_r$ and \textbf{S}$_r$ indicate the wavefronts and Poynting vectors of the negatively reflected HPhPs~\cite{AlvarezPerez22}. \textbf{(e)} Real part of the permittivities of $\alpha$-V$_2$O$_5$ (continuous lines) and $\alpha$'-(Na)V$_2$O$_5$ (dashed lines) extracted from ab-initio calculations along the principal x, y and z axes (red, blue and green lines, respectively). The Reststrahlen bands RB$_{1-3}$, and RB’$_{1-3}$ for $\alpha$-V$_2$O$_5$ and $\alpha$'-(Na)V$_2$O$_5$, are indicated in bright and dark shading, respectively. Green shaded regions represent RB$_1$ and RB’$_1$; red shaded regions represent RB$_2$ and RB’$_2$ and grey shaded regions represent RB$_3$ and RB’$_3$. \textbf{(f)} Experimental near-field images of polariton amplitude $s_3$ in a graphene/$\alpha$-MoO$_3$ heterostructure with graphene doping $E_F=0$ eV (top panel), and $E_F=0.7$ eV (bottom panel), showing hyperbolic and elliptical polariton propagation respectively. Polaritons were launched by a gold antenna. The $\alpha$-MoO$_3$ film was placed on top of a 300 nm SiO\textsubscript{2}/500 $\mu$m Si substrate~\cite{hu2022doping}. \textbf{(g)} Experimental near-field images illustrating negative refraction from a hyperbolic wave in $\alpha$-MoO$_3$ (launched from a gold antenna, see sketch on the right) to an elliptic wave in graphene/$\alpha$-MoO$_3$~\cite{hu2022doping}.}
    \label{fig:fig5}
\end{figure}

Biaxial anisotropy has drawn considerable attention in recent years as it can give rise to highly directional propagation of polaritons along the surface of a material, thereby enabling the direct observation of exotic propagation regimes using conventional microscopy or spectroscopy techniques. Different combinations of permittivity component signs give rise to distinct propagation regimes along the different in-plane directions, as well as different wave characteristics along the surface normal. 
For instance, when both in-plane components are negative, PhPs exhibit in-plane elliptic dispersion, and they can be either confined to the surface, if the out-of-plane permittivity component is also negative, or propagate into the bulk of the material, if it is positive. Notably, when the two in-plane components possess opposite signs, polaritons with in-plane hyperbolic dispersion can propagate along the crystal's surface.

In-plane HPhPs were first reported in $\alpha$-MoO$_3$, an orthorhombic VdW semiconductor with four molecules per unit cell~\cite{Py77,Seguin95} (16 atoms, of which 4 Molybdenum, and 12 Oxygen atoms). As a result of the different bond lengths between the atoms along the three directions (see Ref.~\citeonline{Ratnaparkhe21} for further details), $\alpha$-MoO$3$ exhibits $B_{1u}$, $B_{2u}$, and $B_{3u}$ modes, that are infrared-active and show a TO-LO splitting for electric fields along $z=c$, $y=b$, and $x=a$, respectively. These anisotropic optical phonons along the principal axes of $\alpha$-MoO$_3$ give rise to several spectrally overlapping RBs in the mid-infrared~\cite{AlvarezPerez20} and far-infrared~\cite{deOliveira21} spectral regions (see Fig.~\ref{fig:fig1}). Due to this anisotropy, PhPs in $\alpha$-MoO$3$ can propagate with either elliptic or hyperbolic in-plane dispersion.

In-plane HPhPs were first observed in $\alpha$-MoO$_3$ in the mid-infrared regime using near-field~\cite{ma2018plane} and photo-induced force microscopy~\cite{Zheng19}, as well as nano-FTIR spectroscopy~\cite{ma2018plane}. Fig.~\ref{fig:fig5}a shows a near-field image of HPhPs propagating in an $\alpha$-MoO$_3$ disk, nanoimaged by s-SNOM~\cite{ma2018plane}. HPhPs launched by the tip reflect at the disk edges, travelling back to the tip. The interference pattern shows an almond shape in the hyperbolic RB\textsubscript{2} of $\alpha$-MoO$_3$ (821 - 963 cm\textsuperscript{-1}), with no discernible polariton propagation along the [001] direction, which is thus labeled as a forbidden  direction for propagation of in-plane HPhPs in this spectral range. By Fourier transforming these patterns, the IFC can be directly obtained, clearly showing the in-plane hyperbolic dispersion of HPhPs in this frequency range.

Shortly after the discovery of in-plane HPhPs in $\alpha$-MoO$_3$, HPhPs were also visualized in the far-infrared regime using a free electron laser (FEL)~\cite{deOliveira21}, providing the first report of terahertz HPhPs with nanoscale wavelengths. Fig.~\ref{fig:fig5}b shows the dispersion of PhPs in two different far-infrared RBs of $\alpha$-MoO$_3$. Experimental points were extracted by taking line profiles in s-SNOM images while continuous lines denote the simulated polariton dispersion. Here, the quasi-static approximation $|k|\gg k_0$ was used, in generalizing Eq.~\eqref{eq:disp_ultrathin_hBN} to the case of biaxial slabs~\cite{AlvarezPerez19}, which yields:
\begin{equation}
    k_\parallel(\omega) = \frac{\rho(\omega)}{h}\left\lbrace\text{arctan}\left[\frac{\varepsilon_1}{\varepsilon_z(\omega)\rho(\omega)}\right]+\text{arctan}\left[\frac{\varepsilon_2}{\varepsilon_z(\omega) \rho(\omega)}\right]+\pi l \right\rbrace, \quad  l = 0,1,2..,
    \label{eq:disp_highk_moo3}
\end{equation}
where $\rho(\omega) = {i\sqrt{\varepsilon_z(\omega)}}/{\sqrt{\varepsilon_x(\omega)\text{cos}^2(\phi)+\varepsilon_y(\omega)\text{sin}^2(\phi)}}$, $h$ is the flake thickness and $\phi$ is the angle that the wavevector forms with the x-axis.
Fig.~\ref{fig:fig5}b showcases the strong dependence of the PhP wavelength on the slab thickness, already outlined in the previous section. Thickness variations as small as a few tens of nanometers can give rise to dramatic variations in the confinement, which can be as tight as $\lambda_0/75$, as demonstrated for PhPs in $\alpha$-MoO$_3$ in the far-infrared~\cite{deOliveira21}. This phenomenon can be attributed to the highly dispersive nature of PhPs, which arises from the narrowness of the RB, leading to a remarkably flat dispersion curve and, as a result, very low group velocities of up to $10^{-4}c$, with $c$ the speed of light in free space~\cite{ma2018plane}. These ultra-slow group velocities are experimentally accessible because of the long PhP lifetimes, which are also related to the PhP propagation length, typically of several wavelengths in dielectrics in comparison to metals, where plasmon polaritons are severely damped by their scattering with defects, phonons and quasi-free electrons. Remarkably, $\alpha$-MoO$_3$ supports PhPs with exceptionally long lifetimes, reaching up to 8 ps~\cite{ma2018plane} in RB$_3$, surpassing graphene plasmons at room and even cryogenic~\cite{Ni18} temperatures, as well as isotopically enriched hBN~\cite{giles2018ultralow}. Although the propagation lengths are comparable to those in hBN, the narrowness of this RB in $\alpha$-MoO$_3$ leads to further reduced group velocities. Similar to hBN, the lifetimes and propagation lengths of PhPs in $\alpha$-MoO$_3$ can be increased via isotope enrichment~\cite{Zhao22}, resulting in lifetimes 
as long as $\approx 13.9$ ps, further enhanced at liquid-nitrogen temperatures~\cite{Ni21}. Long lifetimes are particularly desirable for applications that require long-lived light-matter interactions, such as sensing and strong coupling.

The exotic properties of in-plane HPhPs have remarkable consequences at an interface between different polaritonic media. As introduced in the preceding section, the orientation of the Poynting vector and the wavevector (and hence phase velocity) of a hyperbolic wave are strongly direction-dependent and can differ dramatically near the asymptote of the hyperbolic IFC. Due to this non-collinearity, HPhPs can be transmitted or reflected in unconventional directions, since refraction and reflection phenomena require momentum conservation at the interface. Notably, hyperbolic polaritons can deviate away from the normal to the boundary when passing from a medium with a lower refractive index to one with a higher refractive index, in contrast to refraction between isotropic media, where light bends towards the normal under similar conditions. This was demonstrated by near-field imaging of planar nanoprisms milled from SiO\textsubscript{2} substrates underneath $\alpha$-MoO$_3$ flakes~\cite{Duan21}, and has been effectively harnessed for the development of an in-plane refractive hyperlens~\cite{Duan21}. The capability of refracting polaritonic states close to a hyperbolic asymptote offers sub-diffractional resolution compared not only to the wavelength of light in free space ($\lambda_0/50$), but also in relation to the PhP wavelength, resulting in foci as small as $\lambda_p/6$ (where $\lambda_p$ is the polariton wavelength), as shown in Fig. \ref{fig:fig5}c. Foci as small as $\lambda_0/50$ were also demonstrated by using gold nanoantennas of different shapes\cite{MartinSanchez21,Zheng22,Qu22}, that allow effective launching of in-plane HPhPs. In this configuration, states close to the directions given by the hyperbolic IFC dominate when in-plane HPhPs interfere, due to the high density of electromagnetic states along these directions, thereby forming a deeply sub-diffractional focus. 

On the other hand, the non-collinearity between the wavevector and the power flow can also cause polaritons to reflect towards a negative angle, i.e., towards the same side of the normal to the boundary on which they impinge. This was demonstrated by near-field nanoimaging on subwavelength mirrors fabricated in $\alpha$-MoO$_3$~\cite{AlvarezPerez22}, see Fig.~\ref{fig:fig5}d. Both the polaritonic wavelength and direction of propagation are found to depend upon the angle of the boundary and the incident frequency, which can serve as the basis for the design of nanoresonators that exploit the high density of optical modes along the hyperbola asymptotes~\cite{MartinSanchez21,AlvarezPerez22}. The ability to confine in-plane HPhPs to a small volume can dramatically enhance light-matter interactions, reaching Purcell factors and hence facilitating the coupling with quantum emitters. Advancements in plasmonic nanocavity design have enabled the strong coupling of individual emitters to cavity polaritons up to the single-molecule\cite{chikkaraddy2016single} and single quantum dot\cite{Reithmaier04,Leng2018} limit. Nanoresonators made of hyperbolic polaritons can similarly open exciting prospects to translate the fascinating research done on micro-\cite{kavokin2017microcavities}, nano- and picocavities\cite{Benz16} to the realm of phonon polaritons, taking advantage of the high density of electromagnetic modes accompanied by lower losses.

One of the main drawbacks of HPhPs, especially in the context of their technological implementation, is the difficulty in manipulating their in-plane orientation within a device. To face this challenge, 
several methods have been devised to tune the propagation characteristics of HPhPs. For instance, it is possible to steer HPhPs along such forbidden directions by placing a biaxial slab on top of an isotropic polaritonic substrate~\cite{Duan2021_sic,Zhang21}. For the particular configuration $\alpha$-MoO$_3$/4H-SiC, a combination of i) mode hybridization along the direction where both materials have negative permittivity and ii) dielectric loading where the biaxial slab has positive permittivity, leads to HPhP propagation along the [001] direction\cite{Duan2021_sic}, forbidden for a pristine slab of $\alpha$-MoO$_3$ in that frequency range.

More drastic approaches can be adopted to tune the polaritonic dispersion, such as modifying the crystalline structure of a polar dielectric by intercalation. For instance, $\alpha$-MoO$_3$ can be intercalated with hydrogen in a non-volatile and recoverable way, allowing reversible switching of HPhPs while preserving their long lifetimes~\cite{Wu20}. This can also enable spatially selective switching of PhPs in specific regions. An alternative approach is the intercalation of Na atoms into $\alpha$-V$_2$O$_5$, another (less-studied than $\alpha$-MoO$_3$) orthorhombic VdW crystal~\cite{taboada2020broad}. The permittivity along all three principal axes of $\alpha$-V$_2$O$_5$ is displayed in Fig.~\ref{fig:fig5}e, with the RBs of the pristine crystal labeled as unprimed. Intercalation changes the interlayer spacing between the VdW bonded layers, modifying both TO and LO phonon frequencies, and shifting the RBs by as much as 60\% of their initial width (see primed bands in Fig.~\ref{fig:fig5}e). Despite the impressive spectral shifts produced by this intercalation process, the PhPs still show ultra-low losses (lifetime of $\approx 4$ ps), similar to PhPs in the original non-intercalated $\alpha$-V$_2$O$_5$ crystals (lifetime of around $\approx 6$ ps).

Similarly to how the low electron density of plasmonic VdW materials such as graphene facilitates their large tunability via engineering of their Fermi level, which can be achieved by chemical doping or gate voltage, the ability to dynamically control and tune in-plane HPhPs through electronic is highly desirable for future PhP-based integrated technologies, which suggests the engineering of hybrid heterostructures that combine the high tunability of VdW materials with the low losses and hyperbolic response of polar VdW dielectrics. Recent research on $\alpha$-MoO$_3$-graphene heterostructures has indicated that the dispersion of in-plane HPhPs can be tailored through a combination of mode hybridization and Fermi level engineering in the graphene layer~\cite{AlvarezPerez22_graphene,Bapat22,Ruta22,Hu22}. This can be achieved either by chemically modifying the doping level of graphene~\cite{Hu22}, see Fig.~\ref{fig:fig5}f, or by coating graphene with transition-metal dichalcogenides (TMDs) that are subsequently oxidized into transition-metal oxides~\cite{Ruta22}. This process activates charge transfer based on the differing work functions between transition-metal oxides and graphene~\cite{Ruta22}. The former approach (Fig.~\ref{fig:fig5}f) shows the transition from a closed (top panel) to an open (bottom panel) IFC as a result of the doping-dependent hybridization between graphene plasmons and HPhPs in $\alpha$-MoO$_3$. Upon highly enough doping levels, canalized polariton propagation~\cite{hu2020topological,duan2020twisted} is expected to arise in graphene-$\alpha$-MoO$_3$ structures~\cite{AlvarezPerez22_graphene}.

While these methods allow for modifications of the propagation characteristics via external stimuli, they do not provide dynamic control over the in-plane HPhPs. A recent breakthrough demonstrated active control and dynamic switching of PhPs~\cite{Hu23}. By fabricating samples with a SiO\textsubscript{2}/doped-Si substrate, an electric field can be applied perpendicularly to the graphene layer through an SiO\textsubscript{2} dielectric layer coated on a Si backgate. By altering the Fermi level, negative refraction of HPhPs~\cite{Hu23} is gate-tuned through hybridization with graphene plasmons, as shown in Fig.~\ref{fig:fig5}g. The dynamic control of HPhPs clearly represents a key active direction of investigation, and a crucial stepping stone towards their technological application.

\section*{Monoclinic and triclinic crystal system}
\label{sec:monoclinic}
%\section*{Non-Orthorombic Crystals}
%\label{sec:nonorthorhombic}

%intro of mono/triclinic
Further reducing the crystal symmetry from orthorhombic leads to lattice vectors which are no longer orthogonal to each other. In this case, the directions of the transition dipole moments of electronic and vibrational oscillators are no longer locked to the crystal axes~\cite{schubert2016galliumoxide}. Consequently, the permittivity tensor $\hat\varepsilon$ exhibits non-vanishing off-diagonal components in a coordinate system aligned with the crystal axes. Although it is always possible to rotate the coordinate system such that the real part of $\hat\varepsilon$ is diagonal, since $\Re(\hat\varepsilon)$ must be symmetric, these coordinates depend on frequency, a phenomenon known as axial, or angular dispersion~\cite{born2013principles}. Axial dispersion arises from the frequency dependence of the off-diagonal permittivity terms that couple the different polarization directions of the electric field in the material. Some of the key features of optical response in these materials, such as axial dispersion, are discussed in the ellipsometry literature~\cite{claus1978monoclinicoptics,Petit2013monoclinicopticsreview,schubert2016galliumoxide,schubert2017CdWO4,schubert2004infrared,weiglhofer2003introduction}. Yet, the study of polaritons in monoclinic and triclinic crystals is still in its infancy~\cite{passler2022hyperbolic,hu2023real,matson2023controlling}, while polaritons in triclinic systems~\cite{mosshammer2022RS2triclinic} have not been investigated yet to the best of our knowledge.

In monoclinic crystals only two lattice vectors are non-orthogonal, spanning the monoclinic plane, while the third lattice vector is perpendicular to that plane (see Fig.~\ref{fig:fig1}f). This crystal symmetry results in two (out of six) non-vanishing off-diagonal elements of the permittivity tensor, thus coupling the polarization along any two orthogonal directions within this plane. The polaritonic effects that stem from this shear coupling between polarizations has earned the hyperbolic surface modes supported by monoclinic crystals the name \textit{hyperbolic shear polaritons} (HShPs)~\cite{passler2022hyperbolic}. Firstly, the axial dispersion of $\hat\varepsilon$ results in the rotation of the hyperbola axes of HShPs with frequency, dramatically altering their propagation direction~\cite{passler2022hyperbolic,matson2023controlling} (see Fig.~\ref{fig:fig5}e,f). Interestingly, while the degree of structural asymmetry is fixed for a given crystal structure, arising from the relative alignment of the different oscillators, the strength of axial dispersion strongly depends on frequency~\cite{passler2022hyperbolic}. This rotation of the hyperbola axes of HShPs happens concurrently with a (generally different) rotation of the symmetry axes of the imaginary part of $\hat\varepsilon$, which describes the dissipative processes affecting the HShPs. These two different rotations result in a misalignment between the IFCs describing HShP propagation and the effect of dissipation on the HShP lifetimes, leading to a marked asymmetry in their propagation pattern.

To conceptually understand HShPs, we can construct a minimal model system: let us assume a material with only two non-degenerate phonon resonances in the monoclinic plane. Without loss of generality, we can orient our coordinates such that one of the two phonon coordinates is aligned with the $x$-axis. Conversely, the second phonon mode is rotated by an angle $\alpha$ with respect to the $x$-axis, so that the total phononic contribution to the dielectric permittivity in the monoclinic plane is given by
\begin{equation}
    \hat\varepsilon_p = \hat\varepsilon_{p1} + \hat{R}^{-1}(\alpha)\hat\varepsilon_{p2}\hat{R}(\alpha)
\end{equation}
where $\hat\varepsilon_{pj}$ is the permittivity along the $j^{th}$ phonon axes, and $R(\alpha) = \{\cos(\alpha),-\sin(\alpha);\sin(\alpha),\cos(\alpha) \}$ is a rotation matrix. A more sophisticated model is given in Ref.~\citeonline{hu2023real}, including contributions from multiple RBs with arbitrary orientations.
For our simple illustrative model, assuming an isotropic non-resonant background permittivity yields a frequency-dependent OA, whose orientation with respect to the $x$-axis is given by the angle
\begin{equation}
    \beta(\omega;\alpha) = \frac{1}{2}\arctan{\bigg(\frac{\Re[\varepsilon_{p2}(\omega)]\sin(2\alpha)}{\Re[\varepsilon_{p1}(\omega)]+\Re[\varepsilon_{p2}(\omega)]\cos(2\alpha)}\bigg)},
    \label{eq:axial_dispersion}
\end{equation}
where $\varepsilon_{pj}$ is a scalar dielectric function that describes the in-axis response from the $j^{th}$ individual phonon mode described by Eq.~\ref{eq:phonon_oscillator}. Then,  Eq.~\ref{eq:axial_dispersion} demonstrates that the magnitude of the axial dispersion $\partial \beta/\partial \omega$ strongly depends on $\alpha$, as well as on the spectral detuning and linewidth of the two oscillators, and that axial dispersion vanishes for $\alpha = 0, 90^\circ$. Note that the diagonalization process above does not generally diagonalize both Hermitian (real) part and non-Hermitian (imaginary) parts of the permittivity tensor simultaneously. The axial dispersion described by Eq.~\ref{eq:axial_dispersion} ensures that the hyperbola axes rotate with frequency. Yet, after diagonalizing $\real[\hat{\varepsilon}_p]$, the non-Hermitian part $\hat\varepsilon_p$ retains its off-diagonal components in the rotated frame. As a result, losses are distributed asymmetrically in different hyperbola branches, a second remarkable feature of HShPs. Finally, note that hyperbolic propagation can be expected whenever the eigenvalues of the Hermitian part of the tensor in the plane have opposite signs, i.e., when the two permittivity components elements in the monoclinic plane are oppositely-signed in the rotated frame that diagonalizes the tensor. 

\begin{figure}[t]
    \centering
    \includegraphics[width=\textwidth]{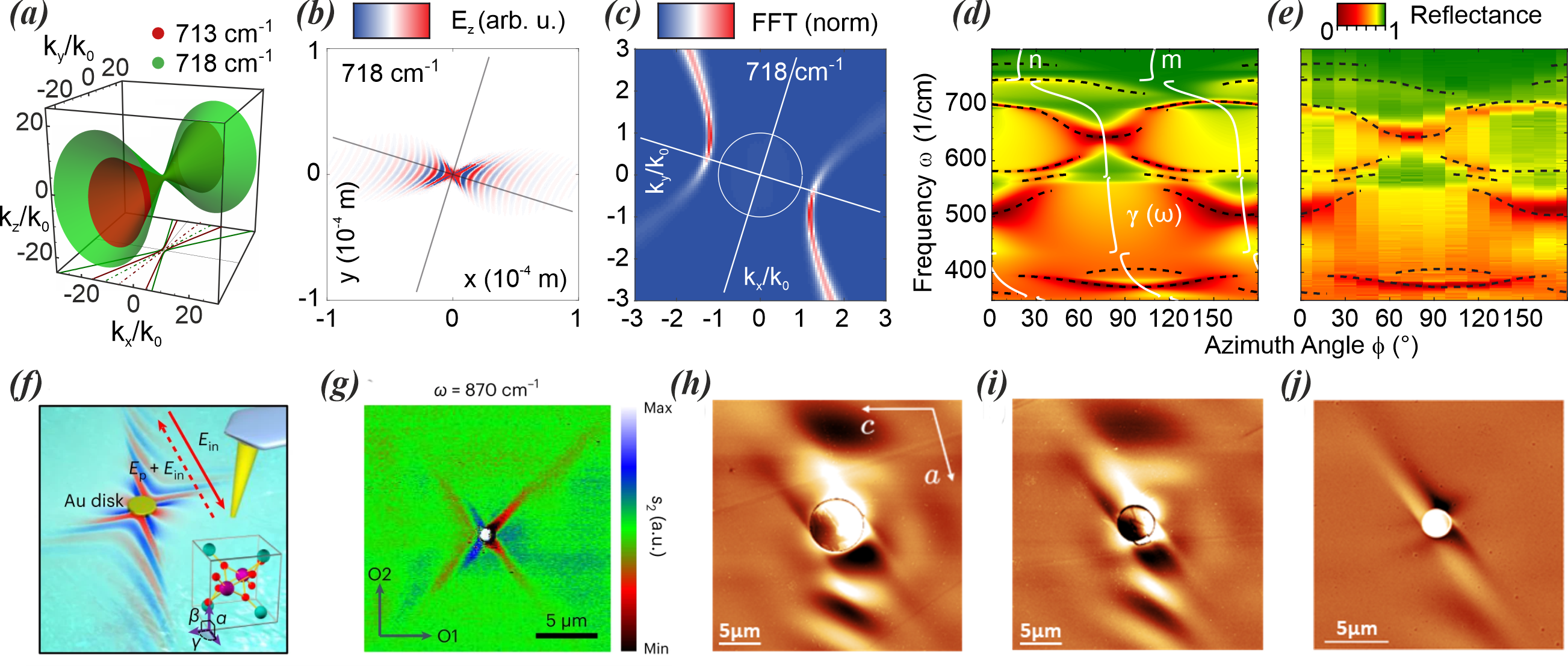}
    \caption{\textbf{Axial dispersion and shear in monoclinic crystals.} \textbf{(a)} Theoretical IFCs (shades) and  their monoclinic surface-plane momentum projections (continuous lines) as well as the in-plane optical axes (dashed lines) for monoclinic $\beta$-Ga$_2$O$_3$ at frequencies $\omega\approx 718$ cm$^{-1}$ (green) and $713$ cm$^{-1}$ (red) clearly showing axial dispersion~\cite{passler2022hyperbolic}. \textbf{(b)} Real and \textbf{(c)} Fourier-space plots of simulated out-of-plane electric field for HShPs on the $\beta$-Ga$_2$O$_3$(010) surface, clearly showing the breaking of mirror-symmetry about the hyperbola axes due to the shear effect~\cite{passler2022hyperbolic}. \textbf{(d)} Simulated and \textbf{(e)} experimental dispersion of HShPs in bGO acquired via Otto-type prism coupling, clearly showing the loss of mirror symmetry in the azimuthal polariton dispersion~\cite{passler2022hyperbolic}, in contrast to higher-symmetry systems. The white lines in \textbf{(d)} show the frequency-dependent OA direction, i.e., the axial dispersion. \textbf{(f)} Illustration of the near-field probing experiment for HShPs in CdWO$_4$: HShPs are launched by an Au disk, and the polaritonic near-field is recorded by a nanotip~\cite{hu2023real}. \textbf{(g)} Real-space image of HShPs in CdWO$_4$ recorded at frequency $\omega=870~\mathrm{cm}^{-1}$~\cite{hu2023real}. \textbf{(h-j)} Real-space images of HShP propagation in $\beta$-Ga$_2$O$_3$ at $\omega=725~\mathrm{cm}^{-1}$, excited with gold antennas of different sizes, demonstrating the enhanced propagation asymmetry exhibited by more confined HShPs~\cite{matson2023controlling}.}
    \label{fig:monoclinic}
\end{figure}

% simulations and first observation
The effect of this OA rotation on hyperbolic bulk polaritons is exemplified in Fig.~\ref{fig:monoclinic}a for $\beta$-Ga$_2$O$_3$, where two IFCs for extraordinary waves in momentum space are plotted, for two closely spaced frequencies $\omega = 713$ cm$^{-1}$ (red) and $\omega = 718$ cm$^{-1}$ (green)~\cite{passler2022hyperbolic}. The projection of the dispersion contour onto the monoclinic $x-y$ plane (continuous lines), along with the respective OA (dashed lines), clearly shows the rotation of the hyperbola with frequency~\cite{schubert2004infrared,passler2022hyperbolic}, as recently confirmed experimentally~\cite{matson2023controlling}. 
In addition, as a consequence of the aforementioned misalignment between the symmetry axes of the real and imaginary parts of $\hat\varepsilon_p$, the conventional mirror-symmetry for polariton propagation is broken, and propagation patterns become strongly skewed, as shown in real- and Fourier-space plots obtained from numerical point-dipole simulations in Fig.~\ref{fig:monoclinic}b and \ref{fig:monoclinic}c respectively, showing the clear asymmetry in the HShP lifetime relative to the OA (perpendicular lines). HShPs were first observed in $\beta$-Ga$_2$O$_3$~\cite{passler2022hyperbolic} via Otto-type prism coupling experiments, and the corresponding simulated and measured polariton spectra are shown in Fig.~\ref{fig:monoclinic}d and \ref{fig:monoclinic}e against frequency (vertical axis) and azimuthal angle (horizontal axis) for a fixed magnitude of the in-plane momentum, clearly evidencing the lack of mirror symmetry in the polariton propagation. Further experiments mapped out the hyperbolic in-plane dispersion, providing the first evidence for polaritonic axial dispersion and propagation asymmetry~\cite{passler2022hyperbolic}.

%Along with this rotation, the propagation of HShPs, shown in real-space in Fig.~\ref{fig:monoclinic}b, becomes extremely skewed~\cite{hu2023real,matson2023controlling}. The 
%It is the difference between real and imaginary part rotation angles that induces the breaking of the symmetry of the propagation patterns~\cite{passler2022hyperbolic}, in stark contrast to higher symmetry systems where this difference vanishes. 
%that stems for the reactive response of the material induced by the interaction between multiple phononic resonances, the dissipative effects in the material do not follow the same symmetry transformation. 

%near-field measurements
In order to rigorously demonstrate the propagation asymmetry of HShPs in real space, direct near-field imaging of the polaritons was achieved soon after for CdWO$_4$~\cite{hu2023real}, and $\beta$-Ga$_2$O$_3$~\cite{matson2023controlling}. These results are reported in Fig.~\ref{fig:monoclinic}f-j, showing (f) the near-field experimental setup and the resulting near-field images for (g) CdWO$_4$ and (h-j) $\beta$-Ga$_2$O$_3$. These data prove the distortion of the polariton field between the two branches of the hyperbola. Furthermore, Fig.~\ref{fig:monoclinic}h-j demonstrate an increase in propagation asymmetry with larger polariton confinement at the interface, achieved by exciting HShPs with gold discs of smaller size, thus coupling to polaritonic modes with increasing momenta~\cite{matson2023controlling}. These studies suggest that the degree of propagation asymmetry of HShPs is not strictly linked to magnitude of the axial dispersion, nor to the degree of structural symmetry breaking. Axial dispersion is weak for CdWO$_4$ compared to $\beta$-Ga$_2$O$_3$, while also the monoclinic angle is much larger for $\beta$-Ga$_2$O$_3$ (103.7$^\circ$) compared to CdWO$_4$ (91.2$^\circ$)~\cite{schubert2017CdWO4,schubert2016galliumoxide}. Yet, both show considerable shear effects on the polariton propagation. The latest finding of the asymmetry increasing with polariton confinement~\cite{matson2023controlling} thus opens vast opportunities for utilizing HShPs for ultra-confined nanophotonic applications.

These first experimental demonstrations of HShPs provide the basis for future applications that will draw from the unique physical and optical phenomena arising in these low-symmetry systems. As a very first example, a giant photonic spin-hall effect was recently predicted\cite{JiaSchubert2023bGOspinhall} to arise from HShPs in $\beta$-Ga$_2$O$_3$. Furthermore, as we will discuss in the following sections, similar optical phenomena could also be observed using specific illumination conditions~\cite{hu2023source} or carefully designed metastructures that specifically harvest the broken symmetry effects that were observed for natural low-symmetry crystals.

\section*{Emergent Directions}

\subsection*{Low-symmetry polaritonics based on twist-optics}
\label{sec:twist-optics}

\begin{figure}
    \centering
    \includegraphics[width=\textwidth]{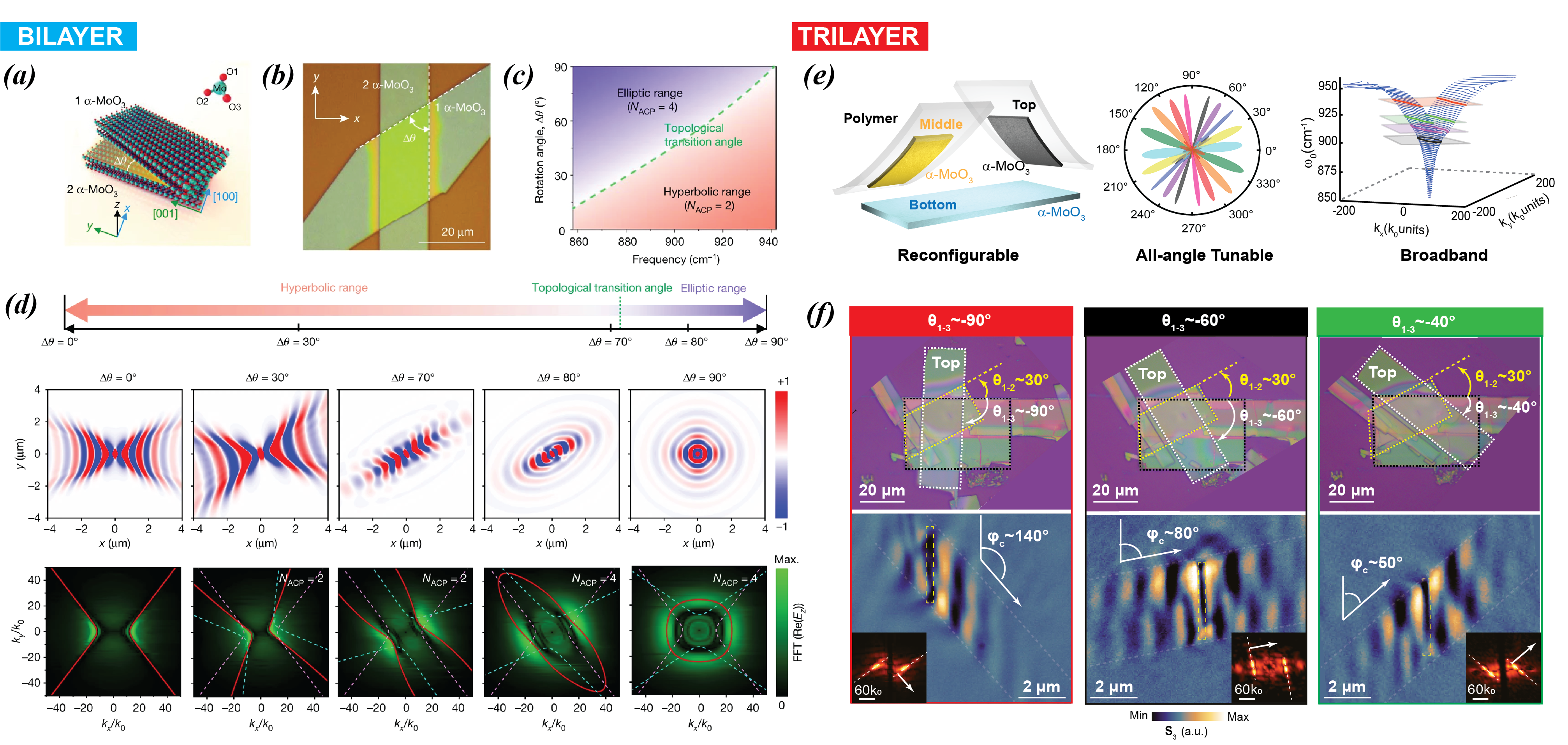}
    \caption{\textbf{Twist-optics: Controlling the propagation of PhPs with twisted bilayer and trilayer VdW stacks.} \textbf{(a-b)} Schematic and optical image of a $\alpha$-MoO$_3$ twisted bilayer stack. \textbf{(c)} Plot of the elliptic-to-hyperbolic topological transition of $\alpha$-MoO$_3$ PhPs as a function of frequency and twist angle. \textbf{(d)} Numerically simulated field distributions (top) and IFCs (bottom) of propagating PhPs at 925.9 cm$^{-1}$ for twisted bilayer stacks with different twist angles $\Delta\theta$. Canalization of PhPs is observed for $\Delta\theta\approx 70^\circ$. \textbf{(e)} Left: Schematic of reconfigurable fabrication of twisted trilayer stacks. Middle: Polar plot of the field intensity in twisted trilayer stacks showing all-angle frequency-tunable polariton canalization. Right: Analytic IFCs of PhPs as a function of incident frequency (colored planes and contours) for twisted trilayer stacks, showing a broadband canalization regime, demonstrated by flattened IFCs in a wide frequency range from 870 to 940 cm$^{-1}$. \textbf{(f)} Top: Optical images of three different twisted trilayer stacks (indicated by angles $\theta_{1-2}$ and $\theta_{1-3}$). All samples are composed of the same three $\alpha$-MoO$_3$ layers. Bottom: s-SNOM near-field amplitude images of the trilayer stacks at 917.4 cm$^{-1}$. The propagation directions of PhPs are marked by white arrows, pointing along different canalization directions $\phi_c=140^\circ$, $\phi_c=80^\circ$, and $\phi_c=50^\circ$. The insets show the corresponding experimental IFCs (FT of the near-field images).}
    \label{fig:fig7}
\end{figure}

The discovery of flat-band unconventional superconductivity~\cite{Cao18a} and Mott insulating behavior~\cite{Cao18b} in twisted graphene bilayers has spurred intense research aiming to uncover novel physical phenomena by twisting layers of VdW materials~\cite{carr2017twistronics}. A prominent example in optics is the superposition of two twisted thin layers of highly anisotropic materials, which provides a neat platform for controlling and steering polariton propagation~\cite{Hu20_moire}. In the spirit of the discussion in the previous sections, the emergent phenomenon here stems from lowering the symmetry of the system by exploiting the in-plane anisotropy of each polaritonic layer, combined with a nontrivial rotation angle between them. As demonstrated by four research groups using near-field nanoimaging on twisted bilayers of $\alpha$-MoO$_3$\cite{hu2020topological,duan2020twisted,chen2020configurable,zebo2020Double} (Fig.~\ref{fig:fig7}a,b), the in-plane polaritonic dispersion undergoes a transition from elliptic to hyperbolic as a function of both frequency and twist angle (Fig.~\ref{fig:fig7}c). In some ways this phenomenon is similar to what was observed in calcite as the angle between OA and interface is rotated (Fig. 3d-e), although it hinges on the twist angle as a symmetry-breaking mechanism. Interestingly, at the transition between the two regimes, the two arms of the polaritonic dispersion become flat and parallel, thereby allowing a single direction for propagation that is diffraction-free, leading to a canalization effect achievable across the entire RB by tuning the twist angle (Fig.~\ref{fig:fig7}d). The twisting of layers to tailor the transition frequency at which the phonon polaritons become canalized provides an innovative avenue to steer and guide light-matter interactions that has since been baptised as \textit{twist-optics}\cite{HerzigSheinfux20}. Once again, tailored broken symmetries provide rational tools to access extreme polaritonic responses. 

Building on this discovery, the effect of a third layer was studied, to explore novel capabilities for controlling the propagation of PhPs\cite{Zheng22_trilayers}. In Ref.~\citeonline{Duan23}, a third layer is repeatedly reassembled (left panel of Fig.~\ref{fig:fig7}e), offering a novel strategy for re-configuring the twist angle in stacked structures. The addition of a third layer provides a major difference with respect to polariton canalization, giving rise to unique features that are not present in twisted bilayers and single layers. For instance, the canalization direction in twisted bilayers is typically limited to a range between 0 and 30 degrees by varying all possible twist angles~\cite{duan2020twisted}. By contrast, with twisted trilayers, the canalization direction can vary between 0 and 180 degrees (middle panel of Fig.~\ref{fig:fig7}e and Fig.~\ref{fig:fig7}f), resulting in all-angle tunable canalization. Moreover, canalization in trilayers can be broadband (right panel of Fig.~\ref{fig:fig7}e), covering a frequency range of up to 70 cm$^{-1}$ (half of the RB under study), and is therefore robust against variations from the magic angles. Finally, IFCs in trilayers can have highly non-trivial (non-intuitive) asymmetric shapes due to the breaking of mirror symmetries, similar to those found in monoclinic crystals.

Just as the electronic version of twistronics has given rise to a flurry of research on superconductivity and exotic electronic states, the photonics variant has important implications for nano-imaging, quantum optics, thermal emission, computing and low-energy optical signal processing.

% [Specially @AA group, please feel free to chip in more stuff here. @Xiang/Andrea: I feel slightly weird that we are only showing simulations about the aMoO3 twist paper, on the other hand there isn't really a clean "theory vs exp" figure do you have any specific suggestions on alternatives here? Also @AP group feel free to make more suggestions (there is, see Extended Data Fig. 7 of ref. hu2020topological]

\subsection*{Low-symmetry metacrystals}
\label{sec:metastructures}

So far, we have only considered naturally occurring anisotropic materials and their heterostructures. The restriction of vibrational resonances to the mid-infrared, however, severely limits the technological impact of low-symmetry crystal polaritonics. Nevertheless, modern fabrication techniques and new insights are in rapid development to artificially realize electromagnetic hyperbolic metamaterials~\cite{lu2012hyperlenses,poddubny2013hyperbolic,ferrari2015hyperbolic,gomez2016flatland,Huo2019Hyper,guo2021zero}, whose emergent response mimics that of low-symmetry polar crystals. A prime advantage offered by such metastructures is the tunability of their frequency response across the electromagnetic spectrum, from the visible~\cite{liu2013metasurfaces,high2015visible} and infrared~\cite{liu2007far,gomez2015hyperbolic,li2018infrared,li2020collective,yang2012experimental}, to radio and microwave frequencies~\cite{chshelokova2012hyperbolic,shchelokova2014magnetic,yermakov2015hybrid,yang2017hyperbolic,yang2018magnetic,yang2019type,yermakov2021surface,chen2021negative,li2022topologically,zheng2019anomalous,Girich2023manipulation}.

\begin{figure}[t]
    \centering
    \includegraphics[width=\textwidth]{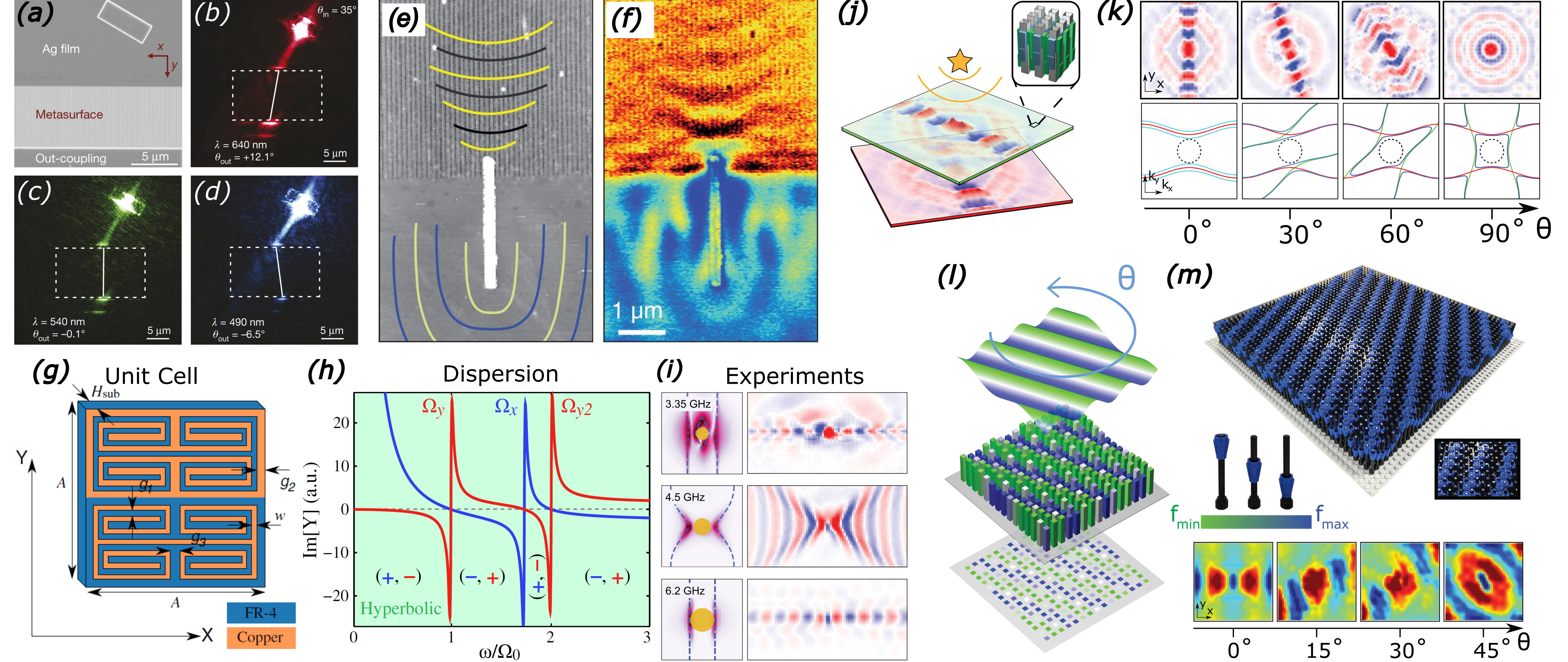} 
    % chose version _v2 for IR->visible->RF->elastic
    \caption{\textbf{Low-symmetry metacrystals}. \textbf{(a)} Nanoscale hyperbolic metasurface realized on a single-crystal silver film supporting surface plasmons in the visible range, enabling \textbf{(b)} positive, \textbf{(c)} null or \textbf{(d)} negative refraction~\cite{high2015visible}. \textbf{(e,f)} In the mid-infrared, a laterally nanostructured hBN monolayer~\cite{li2018infrared} \textbf{(e,top}) provides \textbf{(f,top)} in-plane hyperbolic response by design, in contrast with pristine hBN, which supports conventional in-plane elliptic polaritons (bottom). \textbf{(g-i)} In the microwave  regime, a self-complementary metasurface based on complementary unit cells \textbf{(g)} provides all-frequency hyperbolicity \textbf{(h)}, as clearly demonstrated by electric field maps in the near field \textbf{(i})~\cite{yermakov2021surface}. \textbf{(j,k)} A twisted hyperbolic bilayer metasurface for airborne sound consisting of \textbf{(j)} two layers of strongly coupled tubes which independently support hyperbolic surface waves. The resulting interlayer hybridization enables \textbf{(k)} twist-driven topological transitions from hyperbolic to elliptic behaviour~\cite{yves2022topological}. \textbf{(l,m)} Rotating the anisotropic resonant landscape of an elastic pillared metasurface results in the emergence of moiré patterns \textbf{(l)}, which are responsible for inducing contour-topological transitions, as \textbf{(m)} demonstrated experimentally on a platform made of LEGO$^{\text{TM}}$ blocks~\cite{yves2022moire}.}
    \label{fig:metastructs}
\end{figure}

In this context, nanoscale gratings on a single-crystal silver film, depicted in Fig.~\ref{fig:metastructs}a, act as an on-chip hyperbolic metasurface for surface plasmon polaritons (SPPs) in the visible range~\cite{high2015visible}. Depending on their color, these SPPs undergo either positive, null or negative refraction at the interface with the hyperbolic medium (Fig.~\ref{fig:metastructs}b-d respectively), providing a vivid demonstration of the difference of topology between the corresponding isofrequency contours.

In the mid-infrared, the nano-structuring of an hBN monolayer shown in the top half of Fig.~\ref{fig:metastructs}e makes it effectively behave as an in-plane hyperbolic metasurface, in contrast with the unpatterned portion of the crystal (lower half of Fig.~\ref{fig:metastructs}e), which shows conventional in-plane isotropy~\cite{li2018infrared}. The use of a VdW material is crucial in the realization of such metastructure due to its capability of supporting strongly confined volume polaritons with lower dissipation compared to lossy metal-dielectric metamaterials. On this platform, the resulting polaritons excited by a rod launcher (white bar in Fig.~\ref{fig:metastructs}e) propagate with convex wavefronts in the lower region, where unpatterned hBN trivially behaves as an elliptic medium (bottom of Fig.~\ref{fig:metastructs}f), whereas hyperbolic waves are launched on the metasurface (top of Fig.~\ref{fig:metastructs}f), thereby achieving on-demand hyperbolicity in the propagation of phonon polaritons.

Going down in frequency to the microwave regime, a recent study using complementary unit cells (see Fig.~\ref{fig:metastructs}g) showed that it is possible to construct a self-complementary metasurface with all-frequency hyperbolicity~\cite{yermakov2021surface} (see Fig.~\ref{fig:metastructs}h), thus expanding the capabilities of conventional polaritonic media, where hyperbolicity is limited to the RB. This peculiar property stems from the inherent duality of electromagnetic fields supported by the structure, providing two principal directions along which the metasurface is respectively inductive and capacitive. Self-complementary metasurfaces thus allow one to fully overcome the frequency limitations of conventional surface waves, and to harness the polarization and directionality of hyperbolic waves for an extremely broad range of frequencies (Fig.~\ref{fig:metastructs}i). Hyperbolic metastructures have also attracted considerable interest in the acoustic\cite{li2009experimental,shen2015broadband,ju2018acoustic,quan2019hyperbolic,yves2021extreme} and elastodynamic\cite{oh2014truly,lee2016extreme,zhu2016single,dong2018broadband} domains, enabling the design of polaritonic analogues at larger scales and lower frequency ranges.

These artificial metacrystals also offer an ideal playground to extend the twist-optics paradigm discussed above beyond natural materials. For instance, wave analogs of twisted bilayer graphene, which host flat bands at magic-angles analogous to their electronic counterparts, have been demonstrated both in photonics~\cite{dong2021flat,oudich2021photonic,lou2021theory,lou2022tunable,tang2022chip,yi2022strong} and phononics~\cite{lopez2020flat,deng2020magic,gardezi2021simulating,oudich2022twisted}, yielding twist-reconfigurable resonant behavior which provides novel opportunities for enhanced wave trapping. In the same trend, such tunable moire patterns permit the localization of light~\cite{wang2020localization} and elastic waves~\cite{marti2021dipolar}, quasi-Bound states in the continuum~\cite{huang2022moire}, dynamic beamforming~\cite{liu2022moire} and non-trivial behavior related to the band topology of the twisted crystal~\cite{rosa2021topological,wu2022higher}. Moreover, this twist degree-of-freedom can also be used to induce artificial gauge fields that result in tunable refraction~\cite{cohen2020generalized,yang2021demonstration}. Finally, hyperbolic to elliptic topological transitions have also been implemented in twisted hyperbolic metasurfaces both in photonics~\cite{liu2022magnetic,zheng2022molding,girich2022manipulation} and phononics~\cite{yves2022moire,yves2022topological}.

Fig.~\ref{fig:metastructs}j shows an airborne-acoustic bilayer metasurface, with each layer consisting of a square subwavelength array of tightly coupled waveguides. Each layer hosts hyperbolic surface modes~\cite{yves2022topological}, and as the two metasurfaces are brought within the near-field of each other, their hyperbolic modes hybridize. Interestingly, as shown in Fig.~\ref{fig:metastructs}k, the number of anti-crossing points between the (bottom row) isofrequency contours of the two metasurfaces acts as a topological invariant, dictating a topological transition between a hyperbolic and an elliptic phase of the resulting hybridized surface modes. 

Finally, a similar concept can be realized by embedding the effect of a twist within the extended geometrical structure of a metasurface. In particular, introducing a directional periodic modulation in the properties of the resonators that constitute a metasurface can produce a moiré effect when the periodicity of the lattice structure and that of the resonator modulation are interleaved (Fig.~\ref{fig:metastructs}l). One instance of this paradigm was realized in an elastic metasurface constituted by LEGO$^{\text{TM}}$ pillar resonators, whose resonant frequency can be tuned by varying the center of mass of the pillar as shown in the inset of Fig.~\ref{fig:metastructs}m, along with the experimental sample~\cite{yves2022moire}. Varying the twist angle of such a modulation can also be used to drive topological transitions between hyperbolic and elliptic flexural waves on a thin plate (bottom inset in Fig.~\ref{fig:metastructs}m). Finally on-demand shear hyperbolic flexural waves were recently implemented based on the back-to-back stacking of two elastic pillar metasurfaces, demonstrating complete tunability of the propagation direction of the hyperbolic waves with frequency, full reconfigurability, and giant loss asymmetry, also proposing applications of these concepts for reflectionless negative refraction and defect detection~\cite{yves2023twist}.

\section*{Conclusions and Outlook}
\label{sec:conclusions}

In this Review we have provided a symmetry-driven discussion of the latest developments in the field of phonon polaritonics, highlighting how lowering the crystal symmetry leads to the emergence of new polariton states with increasingly exotic features. These phonon-polaritons can confine electromagnetic fields within deeply subwavelength volumes, while exhibiting much longer lifetimes than plasmon polaritons in metals, thus constituting a truly competitive platform to enhance light-matter interactions. While not yet explored, these features form an ideal basis for nonlinear optics at the nanoscale, in which symmetries may play an even more exciting role. In addition, these polaritons exhibit extremely high directionality, which can be enhanced even further in lower-symmetry platforms such as monoclinic and triclinic crystals. Other polariton families can leak into the bulk of anisotropic crystals, providing a bridge between nanoscale field confinement and far-field radiation, and an interesting trade-off between directionality, confinement, and lifetime. Several applications have been emerging from this platform, ranging from focusing and imaging\cite{caldwell2014sub,dai2015subdiffractional}, to sensing, nanophotonic circuits and routing~\cite{sternbach2023negativerefraction,hu2023gate}.

Remarkably, this paradigm can be further leveraged to achieve on-demand phonon-polaritonic responses by photonic engineering, exemplified by the rising field of twist-optics, whereby layers of anisotropic VdW materials are overlai at a finite twist-angle, broadening the spectrum of opportunities for mid-infrared polaritonics. This concept can also be exported above and below its typical frequency regime by designing artificial low-symmetry metastructures to achieve analogue polaritonic responses across the electromagnetic spectrum, as well as for mechanical waves. This synergy between polaritonics and metamaterials is leading to exciting developments~\cite{zhang2021interface}, which for the most part remain untapped. For instance, topological polaritonic edge states were recently demonstrated in Ref.~\cite{guddala2021topological} by combining a topological photonic crystal with a thin hBN film. The topological crystal was made of a structured array of triangular holes in silicon-on-sapphire. As the size of each unit cell, consisting of six triangular holes, is expanded or shrunken, topologically distinct phases of the lattice are created, capable of hosting topologically protected edge states for light at a domain wall formed by the interface between the expanded and shrunken lattices. As the photonic edge states hybridize with the transverse optical (TO) phonon modes supported by hBN, the resulting polaritonic states inherit the topological order of the photonic edge states. These chiral modes were observed to propagate along sharp edges with high efficiency and propagation lengths of $\approx 80$ $\mu$m, in the direction locked to the specific circular polarization of the impinging waves, thus inheriting the helical nature of these topological edge states. As the wave number of these modes varies, the resulting polaritons have a changing polaritonic nature: for more confined modes, with large wave number, the phonon features are more pronounced, and phonon vibrations acquire topological order. Another possible route to study exotic topological phases consists of polaritonic crystals, consisting of periodic patterns in polaritonic materials. For example, initial experimental attempts were made by fabricating hole arrays in VdW materials and characterizing their ultra-confined Bloch wave features via near-field imaging. The complex polaritonic bands supported by polaritonic crystals were also probed by far-field spectroscopy, enabling new opportunities for exploring topological phases in polaritonic crystals~\cite{alfaro2019deeply,alfaro2021hyperspectral}. Finally, combining chiral structures with VdW crystals is being actively exploited to engineer polaritonic vortices with reconfigurable topological charge~\cite{wang2022spin}.

%Figure \ref{fig:fig8}(f) shows the measured band structure of (left) the bare photonic metasurface, supporting photonic edge states that join the bulk photonic bands at low frequencies, and (right) that of the hybrid metasurface-hBN heterostructure, where the edge polaritonic states are present in the band gap, which instead asymptotically approach the TO phonon line. Characterization of these topological phonon polaritonic modes revealed them carrying a $\approx 50\%$ phononic component, 

On a different note, amidst the ongoing pursuit of temporal photonic control~\cite{galiffi2022photonics,engheta2023four}, several experiments successfully achieved all-optical activation of transient phonon-polaritonic states~\cite{huber2017femtosecond,sternbach2021programmable}. By exploiting the non-equilibrium electron-hole plasma excited by 
%(Fig.~\ref{fig:fig8}(g))
a near-infrared pulse in black phosphorus, transient hybrid plasmon-phonon modes resulting from the interaction between the photoexcited plasma and the inherent vibrational modes of the adjacent SiO$_2$ can be activated and excited via a near-field probe. 
%The intensity of the resulting pump-probe signal, normalized with respect to the measured signal at negative pump-probe delay, is shown in the left panel of Fig.~\ref{fig:fig8}(h), whereas the right panel shows near-field-imaged snapshots of the propagating polariton at different delays, demonstrating its transient nature. 
Similar results were achieved in Ref.~\citeonline{sternbach2021programmable} on layered out-of-plane uniaxial WSe$_2$, where a transition from elliptic to hyberbolic dispersion was induced all-optically, resulting in the onset of transient type-II volume-hyperbolic polaritons based on a positive-to-negative transition of the in-plane components of the dielectric permittivity
%(Fig.~\ref{fig:fig8}(i), left), 
caused by the generation of electron-hole pairs that support excitons. 
%The resulting transient imaging of two nanodisks is shown in the right plot of Fig.~\ref{fig:fig8}(i). 
Thus, temporal control of transient states opens an exciting perspective on the field of phonon polariton photonics. Finally, another exciting direction is constituted by rising field of nonlinear phononics\cite{disa2021engineering,forst2011nonlinear,cartella2018parametric}, and its intersection with polaritonics and nanophotonics promises an entire new avenue of future exploration, particularly as a new playground for Floquet physics~\cite{michael2022generalized,sugiura2022resonantly}, further motivated by the recent observations of phonon-driven transient states of materials~\cite{mitrano2016possible,babadi2017theory,stupakiewicz2021ultrafast}.

To conclude, breaking of symmetries at the nanoscale has been unveiling a treasure trove of opportunities to engineer novel polaritonic and nanophotonic phenomena, many of which still lie undiscovered at the intersection with other exciting fields of research, including time-varying media, exploiting new synergies between spatial and temporal broken symmetries, topological photonics, e.g., leveraging the natural chiral phonon features of monoclinic crystals, and free-electron nanophotonics, among several others. Finally, no studies have yet explored what new polaritonic phenomena may lie untapped in triclinic materials, testifying that there is plenty of room for discovery at the bottom of the symmetry scale.
\bigskip

\noindent\textbf{Acknowledgements} \\
We thank Dr. Christian Carbogno (FHI Berlin) for computing the displacement vectors of the relevant phonon modes shown in Fig. 1. A.A., E.G., X.N., S.Y., E.M.R, R.N. were partially supported by the Simons Foundation, the Air Force Office of Scientific Research with MURI grants No. FA9550-18-1-0379 and FA9550-22-1-0317, and the Office of Naval Research with grant No. N00014-19-1-2011. E.G. acknowledges funding from the Simons Foundation through a Junior Fellowship of the Simons Society of Fellows. G.C., S.W., M.W., and A.P acknowledge support by the Max Planck Society. G.À.-P. acknowledges support through the Severo Ochoa program from the Government of the Principality of Asturias (grant number PA-20-PF-BP19-053). P.A.G. acknowledges support from the European Research Council under Consolidator grant no. 101044461, TWISTOPTICS and the Spanish Ministry of Science and Innovation (State Plan for Scientific and Technical Research and Innovation grant number PID2022-141304NB-I00).
\bigskip

\noindent\textbf{Author contributions} \\
E.G., G.C., and X.N. contributed equally to the article. All authors contributed substantially to the discussion of the content. A.A. initiated the project. E.G., G.C, X.N., G.A.P., P.A.G., S.Y., E.M.R and A.P. researched the data and wrote the respective sections of the article. E.G., G.C., X.N., G.A.P., E.M.R., S.Y., A.P, and A.A. reviewed and edited the manuscript.  

\bibliography{bibtexBiblio}

% \noindent\textbf{Competing interests}\\
% Nature Journals require authors to declare any competing interests in relation to the work described. Information on this policy is available \href{http://www.nature.com/authors/policies/competing.html}{here}. \\

\end{document}